\documentclass[12pt]{iopart}
\usepackage{iopams,epsfig,amsbsy}  

\newcommand{\boldn}{{\mathbf n}}
\newcommand{\boldm}{{\mathbf m}}
\newcommand{\er}{{\mathbb R}}
\newcommand{\en}{{\mathbb N}}

\newcommand{\erf}{\mathop{\mbox{erf}}}

\newcommand{\vol}{\nu}
\newcommand{\dt}{\mbox{d}t}
\newcommand{\dr}{\mbox{d}r}
\newcommand{\imathmy}{\mbox{\bf i}}
\newcommand{\imathmys}{\mbox{\bf \scriptsize i}}
\newcommand{\picturesAB}[4]{
\centerline{
\hskip #4
\raise #3 \hbox{\raise 0.9mm \hbox{(a)}}
\hskip -8mm
\epsfig{file=#1,height=#3}
\raise #3 \hbox{\raise 0.9mm \hbox{(b)}}
\hskip -8mm
\epsfig{file=#2,height=#3}
}}
\newcommand{\picturesABs}[5]{
\centerline{
\hskip #4
\raise #3 \hbox{\raise 0.9mm \hbox{(a)}}
\hskip -8mm
\epsfig{file=#1,height=#3}
\hskip #5
\raise #3 \hbox{\raise 0.9mm \hbox{(b)}}
\hskip -8mm
\epsfig{file=#2,height=#3}
}}
\newcommand{\boldE}{{\mathbf E}}
\newcommand{\bolde}{{\mathbf e}}
\newcommand{\bolddelta}{{\boldsymbol \delta}}
\newcommand{\zet}{{\mathbb Z}}
\newcommand{\dW}{\mbox{d}W}
\newcommand{\newrho}{\overline{\varrho}}
\newcommand{\cn}{\overline{b}}
\newcommand{\tildeM}{\widetilde{M}}

\begin{document}

\title[Stochastic reaction-diffusion processes]{Stochastic modelling 
of reaction-diffusion processes: algorithms for bimolecular reactions}

\author{
Radek Erban and S. Jonathan Chapman
}

\address{University of Oxford, Mathematical Institute,
24-29 St. Giles', Oxford, OX1 3LB, United Kingdom}

\ead{erban@maths.ox.ac.uk; chapman@maths.ox.ac.uk}

\begin{abstract}
Several stochastic simulation algorithms (SSAs) have been recently 
proposed for modelling reaction-diffusion processes in cellular and 
molecular biology. In this paper, two commonly used SSAs are studied.
The first SSA is an on-lattice model described by the reaction-diffusion 
master equation. The second SSA is an off-lattice model based
on the simulation of Brownian motion of individual molecules and 
their reactive collisions. In both cases, it is shown that the commonly
used implementation of bimolecular reactions (i.e. the reactions
of the form $A+B \to C$, or $A + A \to C$) might lead to incorrect
results. Improvements of both SSAs are suggested which overcome
the difficulties highlighted. In particular, a formula
is presented for the smallest possible compartment size (lattice spacing)
which can be correctly implemented in the first model. This 
implementation uses a new formula for the rate of bimolecular 
reactions per compartment (lattice site). 

\end{abstract}




\section{Introduction}

Many cellular and subcellular biological processes can be described 
in terms of diffusing and chemically reacting species (e.g. enzymes) 
\cite{Alberts:2002:MBC,Erban:2007:RBC}. A traditional approach 
to the mathematical modelling of such reaction-diffusion processes is to 
describe each (bio)chemical species by its (spatially-dependent) 
concentration. The time evolution of concentrations is then modelled 
by a system of partial differential equations (PDEs) \cite{Murray:2002:MB}.
Many mathematical and computational methods have been developed over
the last century for solving and analyzing PDEs 
\cite{Thomas:1995:NPD,Zauderer:1983:PDE}, which makes 
PDE-based modelling attractive. However, it has serious limitations
when applied to biological systems. There may be relatively few
numbers of some chemical species; for example, often only one or two 
mRNA molecules of a particular gene are present in the cell
\cite{Alberts:2002:MBC}.
In such cases we cannot even properly define spatially-dependent 
concentration profiles\footnote[7]{The macroscopic concentration
of molecules at a given point in the space is defined as the number
of molecules in a neighbourhood of this point divided by the 
volume of the neighbourhood. In particular, the neighbourhood
must be chosen large enough to contain a lot of molecules. This
is clearly not possible if there are only few molecules present
in the system.}  
and PDE-based models cannot be used. The appropriate quantities
to describe the system are not concentrations, but numbers and
positions of 
molecules of the chemical species involved. 

In recent years, several stochastic simulation algorithms (SSAs)
have been proposed to model the time evolution of molecular 
numbers \cite{Hattne:2005:SRD,Andrews:2004:SSC,Erban:2007:RBC}.
They provide a more detailed and precise picture than
deterministic PDE-based models. They typically give the same results 
for simple systems involving zero-order and first-order chemical 
reactions (for example, linear degradation or conversion). However, 
the situation is more delicate whenever some chemical species are 
subject to bimolecular (second-order) reactions, or
the system under study includes reactive boundaries
(for example, a cellular membrane with receptors).
Reactive boundaries were studied in our previous
paper \cite{Erban:2007:RBC}, where we systematically investigated 
four different SSAs for reaction-diffusion processes which 
had been proposed in the literature. We showed that one 
would obtain incorrect results if the computer implementation of reactive 
boundaries is not handled with care. In particular, what seems
on the face of it the same boundary
condition leads to different results when applied to different 
SSAs. To fix this problem, we derived formulae giving 
the correct relation between experimentally measurable 
characteristics and parameters of the computer implementation 
of boundary conditions for all four SSAs \cite{Erban:2007:RBC}.
A generalization of one of these formulae to anisotropic
diffusion tensors was recently given in \cite{Singer:2008:PRD}.

In this paper, we focus on modelling bimolecular reactions,
i.e. chemical reactions of the form $A + B \to C$
or $A + A \to C$. We investigate two commonly used 
reaction-diffusion SSAs which have been previously implemented
in reaction-diffusion software packages 
MesoRD \cite{Hattne:2005:SRD} and Smoldyn \cite{Andrews:2004:SSC}. 
The first reaction-diffusion SSA is based on dividing the computational 
domain into artificially well-mixed compartments and postulating that only 
molecules which are within the same compartment can react. Diffusion 
is then modelled as jumps between the neighbouring compartments. 
This approach can be mathematically described by the reaction-diffusion
master equation \cite{Isaacson:2006:IDC,Engblom:2008:SSR,Erban:2007:PGS}
and was recently implemented in the mesoscopic reaction-diffusion
simulator MesoRD \cite{Hattne:2005:SRD}. In order to use
this method, we have to choose an appropriate compartment size.
On one hand, the compartment size must be chosen small enough 
so that the spatial variation in the concentration profiles can be 
captured with a desired resolution. The situation is analogous 
to solving  PDEs numerically by a finite difference method.
In order to solve PDEs with the desired accuracy, we need
to choose a sufficiently fine mesh for discretization. On the other 
hand, we will see in Section \ref{secDisComp} that the compartment 
size cannot be chosen arbitrarily small. The analogy with PDEs fails 
here. Unlike in the case of PDEs (for which we get a more accurate 
solution by using a finer discretization), there is a limit on the 
compartment size from below. In Section \ref{secDisComp}, we will show 
that the error of the computation increases as the compartment size 
decreases. In Section \ref{secImpcomp}, we present the formula for the 
smallest compartment size (which can be simulated by this approach) 
and propose an improved SSA which minimizes the simulation error, by 
modifying the reaction rate per compartment. 

The second SSA studied in this paper is based on Brownian motion of 
individual molecules. In its classical formulation
\cite{Smoluchowski:1917:VMT}, it is postulated that two molecules 
(which are subject to a bimolecular reaction) react whenever they 
are within a specified distance (reaction radius) from each other. 
A variant of this method was recently implemented in the software 
package Smoldyn \cite{Andrews:2004:SSC}. One disadvantage of this 
approach is that the reaction radius is, for typical values of the 
bimolecular rate constant and diffusion coefficient, 
unrealistically small compared to the size of individual molecules. 
In Section \ref{secimpsmoland}, we propose
an improved SSA to overcome this difficulty.
It is based on the assumption that two molecules react with the rate
$\lambda$ whenever they are within the distance $\newrho$. The
formula relating $\lambda$, $\newrho$ and the simulation time step
with the experimentally measurable reaction rate constant is derived.
This formula is used for developing a more realistic SSA
for reaction-diffusion processes. 

The paper is organized as follows. In Section \ref{secmodelproblem},
we present illustrative examples which will be used to demonstrate
the results of the paper. In Section \ref{secDis}, we 
present both reaction-diffusion SSAs and summarize their major 
disadvantages. In Section \ref{secImpr}, we present modified
algorithms which are able to overcome the problems 
highlighted in Section \ref{secDis}. To make this paper accessible
to non-mathematicians, Section \ref{secDis} only contains 
the description of improved algorithms and formulae, together
with the results of illustrative computations. The mathematical derivation 
of the formulae presented and the justification of the modified algorithms
are given in Appendices. We finish with a discussion and conclusions 
in Section \ref{secdiscussion}. 

\section{Bimolecular reactions - two model problems} 
 
\label{secmodelproblem} 

A bimolecular reaction is a chemical reaction involving two reacting
molecules. Examples include
$$
A + B \; \mathop{\longrightarrow}^{k} \;\, C + D,
\qquad\qquad 
A + A \; \mathop{\longrightarrow}^{k} \;\, C,
\qquad \mbox{or} \qquad 
A + B \; \mathop{\longrightarrow}^{k} \;\, B,
$$
where the capital letters $A$, $B$, $C$ stand for chemical species
and $k$ is the reaction rate constant, expressed in units of 
volume over time. From the modelling point of view, it
is useful to divide bimolecular reactions into two 
classes, heteroreactions and homoreactions. The term {\it heteroreaction}
will be used for the bimolecular reaction between molecules of two different 
chemical species (for example, heterodimerization $A + B \to C$ 
or catalytic degradation $A + B \to B$). The bimolecular reaction 
between two molecules of the same chemical species 
(for example, homodimerization $A + A \to C$)
will be called the {\it homoreaction} in what follows.
In this section, we introduce two simple chemical systems which 
will be used to illustrate the results in the paper. The first
model will include a heteroreaction (catalytic degradation)
and the second model a homoreaction (homodimerization). More 
complicated examples are discussed later in Section 
\ref{secdiscussion}.

\subsection{A heteroreaction example}

Let us consider chemical species $A$ and $B$ in a container of volume 
$\vol$ which are subject to the following two chemical reactions
\begin{equation}
A + B \; \mathop{\longrightarrow}^{k_1} \;\, B,
\qquad \qquad
\emptyset \; \mathop{\longrightarrow}^{k_2} \;\, A.
\label{abexample}
\end{equation}
The first reaction is the degradation of $A$ catalyzed by $B$. 
We couple it with the second reaction which represents the 
production of molecules of $A$ with the rate\footnote[7]{Let us
note that $k_2$ (the rate constant of the zero-order reaction)
has physical dimension of units per volume per time.
Consequently, $k_2 \vol$ is expressed in units per time.}  
$k_2 \vol$. Since
the number of $B$ molecules is preserved in the chemical 
reactions (\ref{abexample}), the dynamics of the model
(\ref{abexample}) is simple: some molecules of $A$ are produced
by the second reaction and some are destroyed by the first 
reaction. Thus, after an initial transient behaviour, the number
of $A$ molecules fluctuates around its equilibrium value.

Let us consider first the case when the chemical system 
(\ref{abexample}) is well-stirred. Then the probability of 
an ocurrence of the bimolecular reaction is proportial to 
the number of available pairs of reactants.
Let us define the propensity functions of chemical reactions 
(\ref{abexample}) by
\begin{equation}
\alpha_1(t) = A(t) B(t) k_1/\vol,
\qquad 
\alpha_2(t) = k_2 \vol
\label{propabexample}
\end{equation}
where $A(t)$ is the number of molecules of $A$ at time $t$,
$B(t)$ is the number of molecules of $B$ at time $t$, and 
$\vol$ is the system volume. Then the probability that the $i$-th
reaction occurs in the infinitesimally small time interval
$[t,t+\dt)$ is equal to $\alpha_i(t) \, \dt$, $i = 1,2.$
Note that any heteroreaction $A + B \to \cdots$
has a propensity function equal to $\alpha_1(t)$, while the 
propensity function of homoreactions differs; this is the
reason why we discuss them separately.

\begin{figure}
\picturesAB{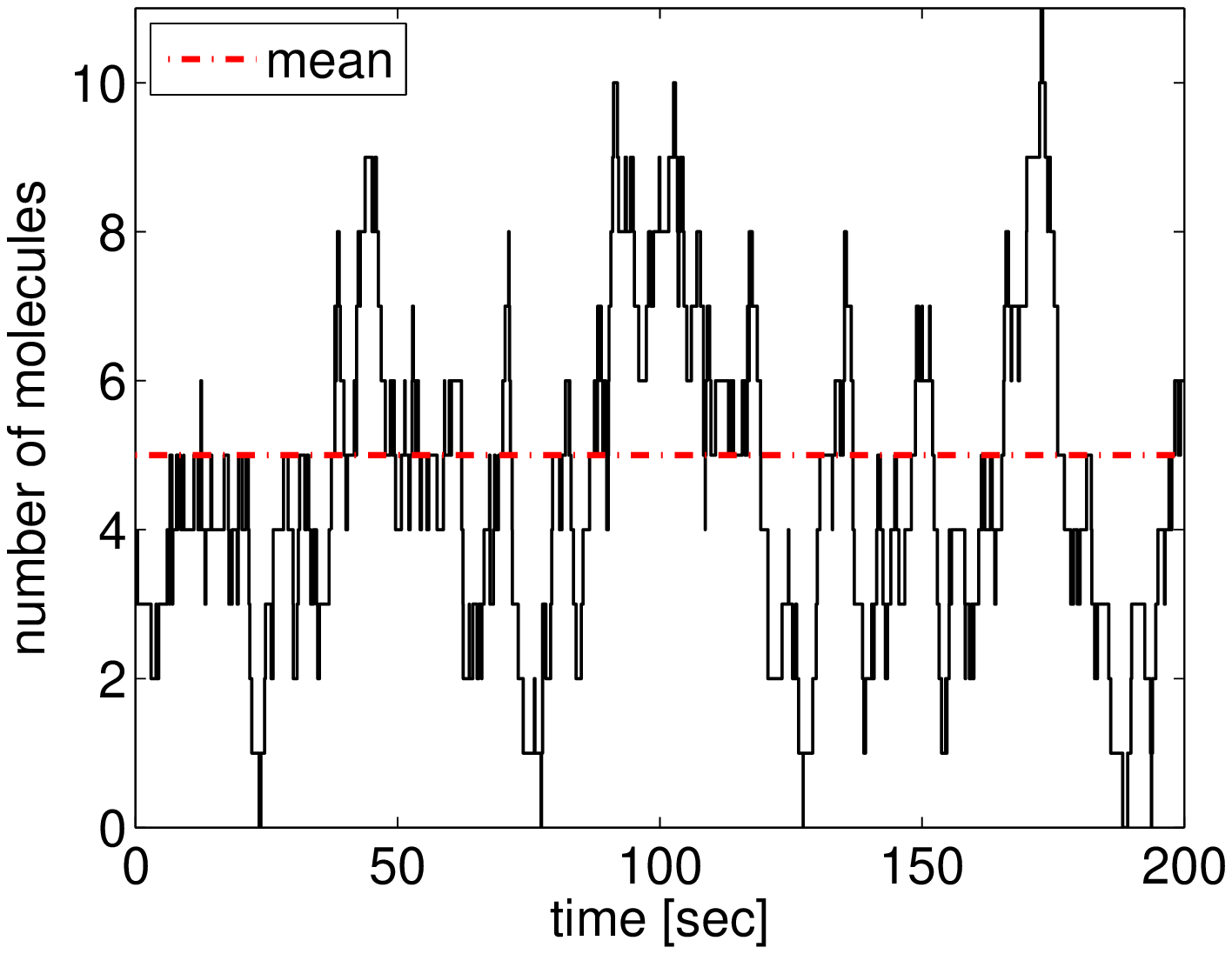}{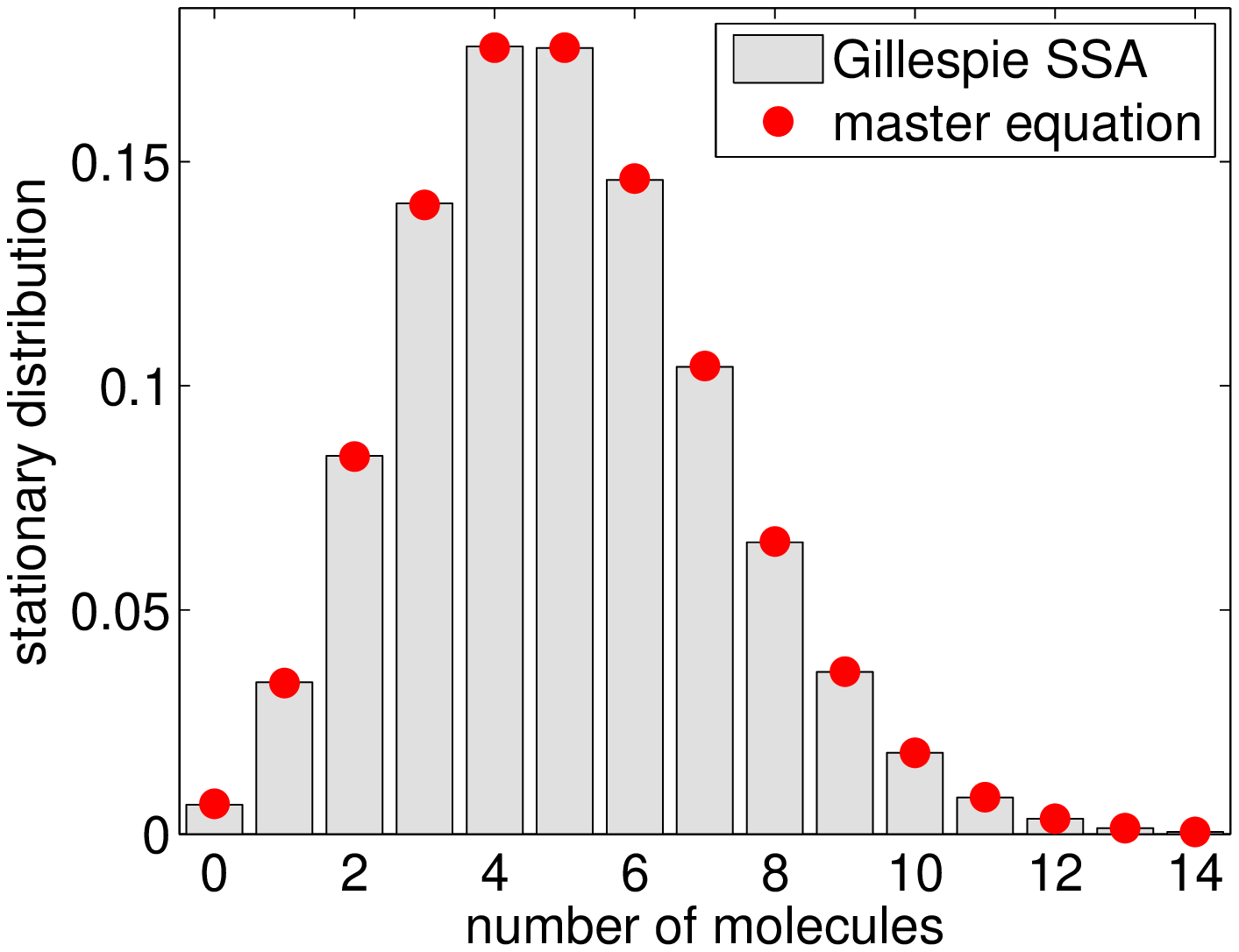}
{2.45in}{5mm}
\caption{{\it Stochastic simulation of the system of chemical
reactions $(\ref{abexample})$ for $k_1/\vol = 0.2 \, \mbox{{\rm sec}}^{-1}$, 
$k_2 \vol= 1 \, \mbox{{\rm sec}}^{-1}$ and one molecule of $B$
in the system. {\rm (a)} $A(t)$ given by one realization 
of the Gillespie SSA (solid line) for the initial value
$A(0)=5$. The average value of $A$ is plotted as the dashed line. 
{\rm (b)} Stationary distribution $\phi(n)$ obtained by long time 
simulation of the Gillespie SSA (grey histogram) and by
formula $(\ref{phindistrabexample})$ (circles).}}
\label{figabexample}
\end{figure}

The chance of occurrence of each reaction is given by the corresponding
propensity function (\ref{propabexample}). If the first chemical 
reaction occurs, then one molecule of $A$ is removed from the system; 
if the 
second chemical reaction takes places, then one molecule of $A$ is 
added to the system. Given the values of the rate constants 
and the initial numbers of molecules of $A$ and $B$, the
stochastic model of (\ref{abexample}) is uniquely specified and 
can be simulated by the Gillespie SSA
\cite{Gillespie:1977:ESS,Gillespie:1992:MPI}. 
In Figure \ref{figabexample}(a), we plot $A(t)$, computed by 
the Gillespie SSA, as the solid line, using
parameter values $k_1 /\vol = 0.2 \, \mbox{sec}^{-1}$, 
$\; k_2 \vol = 1 \, \mbox{sec}^{-1}$, $\; A(0)=5$ and 
$B(0)=1$.
As expected the number of molecules of $A$ fluctuates around
the average value which is $5$ for our parameters.
The nature of these fluctuations can be summarized in terms 
of the {\it stationary distribution}. To compute it,
we record the values of $A(t)$ at equal time intervals and 
create a histogram of the recorded values. Dividing the histogram 
by the total number of recordings, we obtain distribution 
$\phi(n)$ which is plotted in Figure \ref{figabexample}(b)
as the grey histogram. Thus, $\phi(n)$ is the probability
that there are $n$ molecules of $A$ in the system, provided
that the system is observed for a sufficiently long time.
Note that since the Gillespie SSA makes
use of random numbers to compute the time evolution of the 
system, the computed $A(t)$ depends on a particular
realization of the algorithm. Repeating the computation
(with a different set of random numbers), we will obtain
a different time evolution than the one plotted in Figure 
\ref{figabexample}(a). However, the stationary distribution
$\phi(n)$ is uniquely determined by the values of the
rate constants $k_1$, $k_2$ and the number of molecules of $B$ 
in the container. It can be shown
(see \ref{secmathanalmod}) that $\phi(n)$ is the Poisson distribution
\begin{equation}
\phi(n) = \frac{1}{n!} \left( \frac{k_2 \, \vol^2}{k_1 \, B_0} \right)^{\!n} 
\exp\left[ - \frac{k_2 \, \vol^2}{k_1 \, B_0} \right]
\label{phindistrabexample}
\end{equation}
where $B_0$ is the (constant) number of molecules of $B$ in the 
container. The results given by the formula (\ref{phindistrabexample})
are plotted in Figure \ref{figabexample}(b) as the circles.
We confirm that the stationary distribution $\phi(n)$ is indeed
given by (\ref{phindistrabexample}). In what follows, we will use the
stationary distributions of model problems to study the limitations
of different stochastic reaction-diffusion methods. 
The average number of molecules of $A$ in the container is given
by (see \ref{secmathanalmod})
\begin{equation}
M_s 
=
\frac{k_2 \, \vol^2}{k_1 \, B_0}.
\label{meanMsabexample}
\end{equation}
Using the parameter values of Figure \ref{figabexample}, we obtain
$M_s = 5.$ This number is plotted in Figure \ref{figabexample}(a)
at the dashed line. The variance of the Poisson distribution
(\ref{phindistrabexample}) is equal to its mean $M_s$.

\subsection{A homoreaction example}

\label{sechomoreactionexample}

Let us consider chemical species $A$ and $B$ in a container of volume 
$\vol$ which are subject to the following two chemical reactions
\begin{equation}
A + A \; \mathop{\longrightarrow}^{k_1} \;\, B,
\qquad \qquad
\emptyset \; \mathop{\longrightarrow}^{k_2} \;\, A.
\label{nonlindegradationcreation}
\end{equation}
The first reaction describes the dimerization of the chemical
$A$ with the rate constant $k_1$.  We couple it with the second 
reaction which represents the production of the chemical $A$ with 
the rate constant $k_2$. This reaction has been already studied
in the example (\ref{abexample}). In what follows,
we will only be interested in the time evolution and stationary
distribution of $A$. The dynamics of the model
(\ref{nonlindegradationcreation}) is similar to
(\ref{abexample}): some molecules of $A$ are produced
by the second reaction and some are removed by the first 
reaction. Thus, $A(t)$ fluctuates around its equilibrium value
in a similar way as the trajectory in Figure \ref{figabexample} 
which has been computed for the heteroreaction example 
(\ref{abexample}). If the reactor is well-stirred, the propensity 
functions of chemical reactions (\ref{nonlindegradationcreation}) 
are given by
\begin{equation}
\alpha_1(t) = A(t) (A(t)-1) k_1/\vol,
\qquad 
\alpha_2(t) = k_2 \vol,
\label{propglo}
\end{equation}
i.e. the probability that the $i$-th reaction in
(\ref{nonlindegradationcreation}) occurs in the infinitesimally 
small time interval $[t,t+\dt)$ is $\alpha_i(t) \, \dt$, $i = 1,2.$
If the first chemical reaction occurs, then two molecules of $A$ 
are removed from the system; if the second chemical reaction takes 
place, then one molecule of $A$ is added to the system. This 
uniquely specifies the stochastic model as a Markov chain 
which can be simulated by the Gillespie SSA. The stationary
distribution of (\ref{nonlindegradationcreation}) 
is given by (see \ref{secmathanalmod})
\begin{equation}
\phi(n) 
= 
\frac{C}{n!} \left( \frac{k_2 \vol^2}{k_1} \right)^{\!n}
I_{n-1} \!\!\left( 2 \sqrt{\frac{k_2 \vol^2}{k_1}} \right),
\qquad n = 0,1,2,3, \dots,
\label{pnmgfstatGs}
\end{equation}
where $I_{n}$ is the modified Bessel 
function of the first kind (see Glossary) and $C$ is
a positive constant given by the normalization 
$\sum_n \phi(n) = 1.$ The average number of molecules of 
$A$ in the container is given by (see \ref{secmathanalmod})
\begin{equation}
M_s
=
\frac{1}{4} 
+
\sqrt{\frac{k_2 \vol^2}{2 k_1}} \, I_1^\prime 
\!\!\left( 2 \sqrt{\frac{2 k_2 \vol^2}{k_1}} \right)
\left[
I_1 \!\!\left( 2 \sqrt{\frac{2 k_2 \vol^2}{k_1}} \right)
\right]^{\!-1}.
\label{formulaMsGs}
\end{equation}
In \ref{secmathanalmod}, we show that the stationary
number of molecules of $A$ obtained
by the standard ordinary differential equation (ODE) model  
of the chemical system (\ref{nonlindegradationcreation}) is
$\overline{A}_s = \vol \sqrt{k_2/2 k_1}$. Note that this is 
not in general equal to $M_s$ given by formula (\ref{formulaMsGs}).
For example, in the following section, we 
use the parameter values $k_1 /\vol = 0.2 \, \mbox{sec}^{-1}$ and 
$k_2 \vol = 10 \, \mbox{sec}^{-1}$. 
Then $\overline{A}_s=5$ and $M_s \doteq 5.13$, 
i.e. the deterministic ODE does not provide the exact 
description of the stochastic mean. On the other hand,
the difference between $M_s$ and $\overline{A}_s$ is only
2.5\% so that the ODE model gives a reasonable approximation of $M_s$. 
However,  the comparison of deterministic 
and stochastic modelling is not the focus of this paper. 
Our main goal is to highlight some limitations of current
reaction-diffusion SSAs and present improvements
of these models. Examples of chemical systems where 
the differences between the results of stochastic simulation 
and the corresponding deterministic approximation (ODEs)
are significant can be found in 
\cite{Paulsson:2000:SFF,Erban:2007:PGS,DeVille:2006:NDL,Erban:2008:ASC}.

\section{Disadvantages of current SSAs for reaction-diffusion modelling}

\label{secDis}

In Section \ref{secmodelproblem}, we considered the illustrative
examples (\ref{abexample}) and (\ref{nonlindegradationcreation}) as 
well-stirred chemical systems. In particular, their mathematical models 
did not explicitly involve a description of molecular diffusion. 
In this section, we couple chemical systems (\ref{abexample}) and 
(\ref{nonlindegradationcreation})
with models of molecular diffusion and show key limitations
of reaction-diffusion SSAs in the literature. In what follows,
we assume that chemical species $A$ and $B$ diffuse with  
diffusion constants $D_A$ and $D_B$, respectively, in the cubic 
container $[0,L] \times [0,L] \times [0,L]$. We consider
zero-flux (reflective) boundary conditions, i.e.
whenever a molecule hits the boundary, it is reflected back.
The implementation of more complicated (reactive) boundary 
conditions was studied in our previous paper \cite{Erban:2007:RBC}.

\subsection{Compartment-based model}

\label{secDisComp}

In the compartment-based model, we divide the computational
domain into small compartments which are assumed to be
well-mixed. We postulate that only molecules in the same
compartment can react according to bimolecular reactions. 
Diffusion is modelled as jumps of molecules between neighbouring 
compartments \cite{Hattne:2005:SRD,Isaacson:2006:IDC}.

Let us consider the heteroreaction example (\ref{abexample}). 
We divide the cubic domain $[0,L] \times [0,L] \times [0,L]$
\begin{figure}
\centerline{\epsfig{file=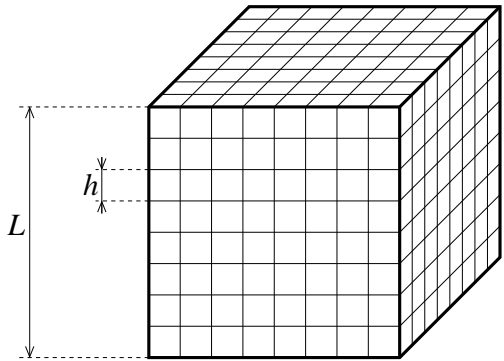,height=2.2in}}
\caption{
{\it Domain $[0,L] \times [0,L] \times [0,L]$
is divided into $K^3$ compartments of the volume
$h^3 = (L/K)^3$. The division of the domain for
$K=8$ is shown on the picture.}
}
\label{figsquare}
\end{figure}
into $K^3$ cubic compartments of volume $h^3$ where $K \ge 1$ and
$h = L/K$ (see Figure \ref{figsquare}).
In general, the compartment-based model can be formulated
for compartments which are not cubic and which are not of the
same size \cite{Isaacson:2006:IDC,Engblom:2008:SSR}. However, for 
the purposes
of this paper, it is sufficient to work with cubic compartments
of the same size. They are the most natural choice, and are easy
to implement computationally. Moreover, if the modeller does
not use the uniform cubic mesh, it might be sometimes difficult
to distinguish which results show a genuine property of the system
and which are a consequence of the non-uniform mesh. Note that the 
uniform cubic mesh introduces an artificial anisotropy in the
domain (e.g. compartments have different lengths along the side
and along the diagonal). However, we will not explore potential
consequences of this anisotropy in this paper. We will focus on 
the more fundamental
problem: the appropriate choice of compartment size $h$.

To precisely formulate the compartment-based SSA for the illustrative
chemical system (\ref{abexample}) in the reactor
$[0,L] \times [0,L] \times [0,L]$, we denote the compartments 
by indices from the set
\begin{equation*}
I_{all}
=
\big\{ (i,j,k) \, | \, i,j,k \; \mbox{are integers such that} \;
                    1 \le i,j,k \le K \big\}.
\end{equation*}
Let $A_{ijk}(t)$ (resp. $B_{ijk}(t)$)  be the number of molecules 
of the chemical species $A$ (resp. $B$) 
in the $(i,j,k)$-th compartment at time $t$
where $(i,j,k) \in I_{all}$. Diffusion is modelled as a jump 
process between neighbouring compartments. Let us define the 
set of possible directions of jumps
\begin{equation*}
\boldE = \{ [1,0,0], \; [-1,0,0], \; [0,1,0], 
            \; [0,-1,0], \; [0,0,1], \; [0,0,-1] \}.
\end{equation*}
For every $(i,j,k) \in I_{all}$, we also define 
\begin{equation*}
\boldE_{ijk}
=
\left\{ \bolde \in \boldE \, | \, 
\big((i,j,k)+\bolde\big) \in I_{all} \right\},
\end{equation*} 
i.e $\boldE_{ijk}$ is the set of possible directions of jumps
from the $(i,j,k)$-th compartment. The compartment-based
reaction-diffusion model can be written using the chemical
reactions formalism as follows. We study a system of $2 K^3$
``chemical species" $A_{ijk}$ and  $B_{ijk}$ for $(i,j,k) \in I_{all}$
which are subject to the chemical reactions:
\begin{equation}
A_{ijk} + B_{ijk} \; \mathop{\longrightarrow}^{k_1} \;\,  B_{ijk},
\qquad \qquad
\emptyset \; \mathop{\longrightarrow}^{k_2} \;\, A_{ijk},
\qquad \mbox{for} \; (i,j,k) \in I_{all},
\label{abexamplecompartments}
\end{equation}
\begin{equation}
A_{ijk} \; \mathop{\longrightarrow}^{D_A/h^2} \;\, A_{ijk+\bolde},
\qquad \mbox{for} \; (i,j,k) \in I_{all}, \; \bolde \in \boldE_{ijk},
\label{diffusionjumpsA}
\end{equation}
\begin{equation}
B_{ijk} \; \mathop{\longrightarrow}^{D_B/h^2} \;\, B_{ijk+\bolde},
\qquad \mbox{for} \; (i,j,k) \in I_{all}, \; \bolde \in \boldE_{ijk}.
\label{diffusionjumpsB}
\end{equation}
The chemical reactions (\ref{abexamplecompartments})
correspond to the chemical system (\ref{abexample})
considered in each compartment. It is assumed that each compartment 
is effectively well-stirred. A molecules of $A$ and a molecule of $B$
which are in the same compartment can react according to the bimolecular 
reaction
$A + B \to B$. On the other hand, two molecules in different 
compartments cannot react with each other. The propensity functions
of reactions  (\ref{abexamplecompartments}) are 
\begin{equation}
\alpha_{ijk,1}(t) = A_{ijk}(t) B_{ijk}(t) \, k_1/h^3, 
\qquad
\alpha_{ijk,2}(t) = k_2 h^3 
\label{propfunc}
\end{equation}
where $h^3$ is the volume of the compartment. The
propensity functions (\ref{propfunc}) can be derived 
using the same argument as (\ref{propabexample}), replacing
the volume $\vol = L^3$ of the whole reactor by the compartment 
volume $h^3$. The reactions (\ref{diffusionjumpsA})--(\ref{diffusionjumpsB})
correspond to diffusive jumps between neighbouring compartments. The
propensity functions of these ``reactions" are equal to
$A_{ijk}(t) \, D_A/h^2$ and $B_{ijk}(t) \, D_B/h^2$. 
There are $2K^3$ reactions in (\ref{abexamplecompartments}),
$6 K^3 - 6 K^2$ diffusion ``reactions" in (\ref{diffusionjumpsA})
and $6 K^3 - 6 K^2$ diffusion ``reactions" in (\ref{diffusionjumpsB})
because there are $6$ possible directions to jump
from each inner compartment and some directions are missing
for boundary compartments. Thus we are able to formulate the
compartment-based reaction-diffusion model as the chemical
system of $2 K^3$ chemical species $A_{ijk}$ and $B_{ijk}$ which 
are subject to $14 K^3 - 12 K^2$ reactions
(\ref{abexamplecompartments})--(\ref{diffusionjumpsB}). 
The time evolution of the chemical system
(\ref{abexamplecompartments})--(\ref{diffusionjumpsB})
can be simulated by the Gillespie SSA. It can be also
equivalently described in terms of the reaction-diffusion 
master equation, which is given in 
\ref{secRDMEmodsys} as equation (\ref{cmeRD}).
The number of molecules of $A$ in the whole container 
$[0,L] \times [0,L] \times [0,L]$ is given by
\begin{equation*}
A(t) = \sum_{(i,j,k) \in I_{all}} A_{ijk}(t).
\label{sumAt}
\end{equation*}
Let $p_n(t)$ be the probability that $A(t) = n.$ Let $\phi_K(n)$
be the stationary distribution defined by (compare
with (\ref{statdistrp}))
\begin{equation}
\phi_K(n) = \lim_{t \to \infty} p_n(t).
\label{statdistrpK}
\end{equation}
Thus, $\phi_K(n)$ is the probability that there are $n$ molecules
of $A$ in the system, provided that the system is observed for
long time. In particular, $\phi_1(n)$ is equal to the stationary 
distribution $\phi(n)$ given by (\ref{phindistrabexample}).
Since production of $A$ is homogeneous throughout the container,
we would expect the distribution of $A$ to be uniform in space,
so that we should find that $\phi_K$ is independent of $K$.
\begin{figure}
\picturesAB{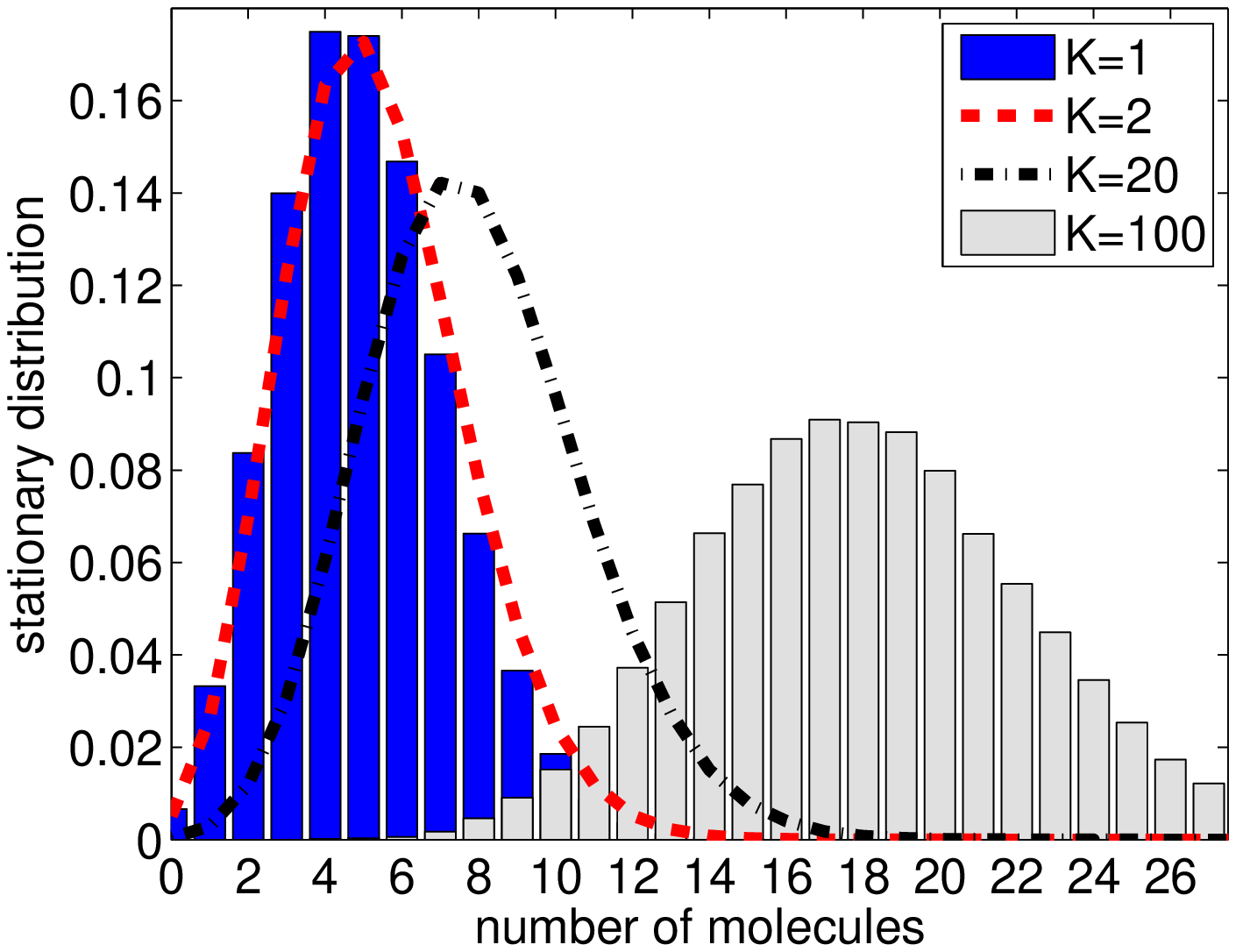}{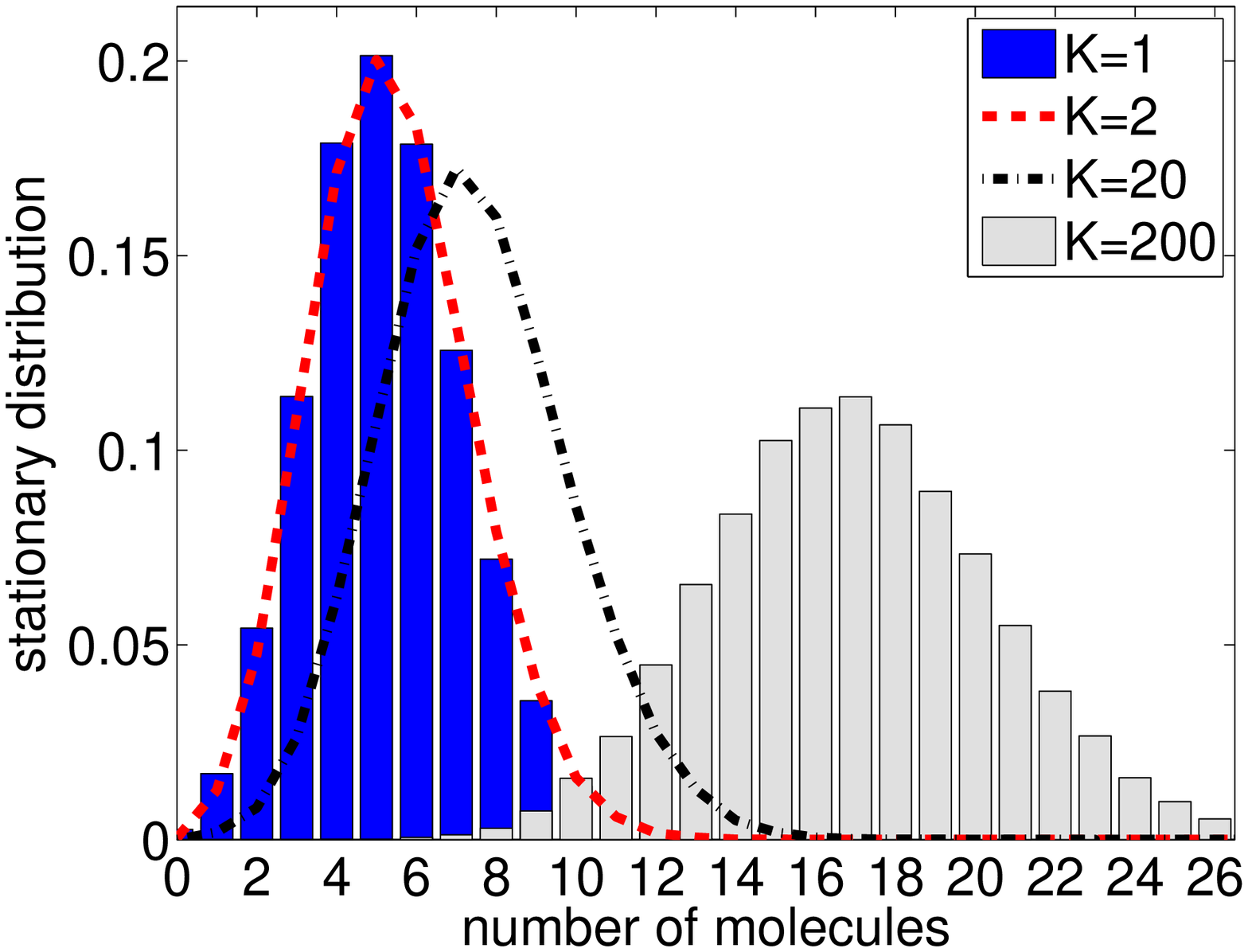}
{2.45in}{5mm}
\caption{(a) {\it Heteroreaction example $(\ref{abexample}).$
Stationary distribution $\phi_K(n)$, 
defined by $(\ref{statdistrpK})$ and computed for 
$K=1, 2, 20$, and $100$ by long time simulations of 
the Gillespie SSA. We use 
$k_1 = 0.2 \; \mu\mbox{{\rm m}}^3 \; \mbox{{\rm sec}}^{-1}$,
$k_2 = 1 \; \mu\mbox{{\rm m}}^{-3} \; \mbox{{\rm sec}}^{-1}$,
$D_A = D_B = 1 \; \mu\mbox{{\rm m}}^{2} \; \mbox{{\rm sec}}^{-1}$,
$L = 1 \; \mu\mbox{{\rm m}}$ and $B_0=1$.}
(b) {\it Homoreaction example $(\ref{nonlindegradationcreation}).$
Stationary distribution $\phi_K(n)$ for 
$K=1, 2, 20$, and $200$ computed by long time simulations 
of the Gillespie SSA. We use 
$k_1 = 0.2 \; \mu\mbox{{\rm m}}^3 \; \mbox{{\rm sec}}^{-1}$,
$k_2 = 10 \; \mu\mbox{{\rm m}}^{-3} \; \mbox{{\rm sec}}^{-1}$,
$D_A = 1 \; \mu\mbox{{\rm m}}^{2} \; \mbox{{\rm sec}}^{-1}$
and $L = 1 \; \mu\mbox{{\rm m}}.$
}}
\label{figlargeKrdme}
\end{figure}
In Figure \ref{figlargeKrdme}(a), we present the stationary 
distributions $\phi_K(n)$ for $K=1, 2, 20$ and $100$
for the parameter values $L = 1 \; \mu\mbox{m}$, 
$D_A = D_B = 1 \; \mu\mbox{m}^2 \, \mbox{sec}^{-1}$,
$k_1 = 0.2 \; \mu\mbox{m}^3  \, \mbox{sec}^{-1}$,
$k_2 = 1 \; \mu\mbox{m}^{-3} \, \mbox{sec}^{-1}$
and $B_0 = 1$.
In particular, we have the same rates for $K=1$ as
were used in Figure \ref{figabexample}, 
namely $k_1 / L^3 = 0.2 \; \mbox{sec}^{-1}$ and 
$k_2 L^3 = 1 \; \mbox{sec}^{-1}$. Thus the 
stationary distribution $\phi_1(n)$, plotted in
Figure \ref{figlargeKrdme}(a), is equal to the
distribution $\phi(n)$ plotted in Figure 
\ref{figabexample}(b).

Increasing  $K$ (i.e. decreasing $h$), the stationary 
distribution $\phi_K(n)$ moves to the right. The shift 
to the right is in agreement with the result of Isaacson 
\cite{Isaacson:2008:RME} who showed
that, in the theoretical limit $h \to 0$, the bimolecular 
reaction $A + B \to \emptyset$ is lost and the 
compartment-based modelling of this reaction only recovers 
the diffusion process. In our case, we coupled 
the bimolecular reaction with the production of $A$ molecules. 
The production rate per the whole domain is equal to 
\begin{equation*}
\sum_{(i,j,k) \in I_{all}} \alpha_{ijk,2}(t)
=
\sum_{(i,j,k) \in I_{all}} k_2 h^3 
=
K^3 k_2 h^3 = k_2,
\end{equation*}
i.e. it is independent of $h$. Thus, for small $h$, 
the slower removal of $A$ by the bimolecular reaction 
and unchanged production rate of $A$ result in the shift of the 
stationary distribution $\phi_K(n)$ to the right in 
Figure \ref{figlargeKrdme}(a), as $K$ is increased 
(i.e. as $h=L/K$ is decreased).
In Figure \ref{figlargeKrdme}(b), we present the results
of a similar computation for the homoreaction example
(\ref{nonlindegradationcreation}). In this case,
the homodimerization $A + A \to B$ is replaced by
$K^3$ reactions $A_{ijk} + A_{ijk} \to B_{ijk}$ for $(i,j,k) \in I_{all}$.
The propensity functions of these reactions are given
by 
\begin{equation}
\alpha_{ijk,1}(t) = A_{ijk}(t) (A_{ijk}(t)-1) \, k_1/h^3.
\label{prophomordme}
\end{equation}
The production reaction and diffusion are treated as in 
(\ref{abexamplecompartments})--(\ref{diffusionjumpsA}).
In Figure \ref{figlargeKrdme}(b), we present the stationary 
distributions $\phi_K(n)$ for four values of $K$.
Notice that $\phi_1(n)$ is equal to the stationary 
distribution $\phi(n)$ given by (\ref{pnmgfstatGs}).
We observe the same phenomenon (shift of the histogram
to the right) as in the case of the heteroreaction
example.

Although it is generally agreed in the literature that there 
is a bound on $h$ from below 
\cite{Isaacson:2006:IDC,Isaacson:2008:RME},
this bound is usually stated in the form $h \gg k_1/(D_A+D_B)$
or $h \gg \varrho$ where $\varrho$ is the binding radius
for the molecular based Smoluchowski model -- see equation
(\ref{reactionradiussmol}) and the discussion in Section \ref{secDisSmol}.
To satisfy these conditions in our particular example,
we could simply choose $h=L$. However, the real importance
of stochastic reaction-diffusion modelling is not in
modelling of spatially homogeneous systems. If the
system has some spatial variations (i.e. some parts
of the computational domain are more preferred by 
molecules than the others), then we obviously want
to choose $h$ small enough to capture the desired spatial
resolution. This leads to the restriction on $h$ from
above, namely $L \gg h$. Thus it is suggested to choose
$h$ small (to satisfy $L \gg h$) but not too small
(to satisfy $h \gg k_1/(D_A+D_B)$) \cite{Isaacson:2006:IDC}
which leads to the important question what the optimal
choice of $h$ should be to get the most accurate results.
In this paper, we propose a different route to this
problem. In Section \ref{secImpcomp}, we show that
there exists a critical value $h_{crit}$ such that
the propensity function of the compartment-based
model can be adjusted for $h \ge h_{crit}$ to
recover correctly the stationary distribution
$\phi(n)$. Thus we effectively replace the 
condition $h \gg k_1/(D_A+D_B)$, which requires that $h$
is much larger than $k_1/(D_A+D_B)$ by a sharp inequality
$h \ge h_{crit}$, where $h_{crit}$ is approximately
a quarter of $k_1/(D_A+D_B)$. We will show that the 
compartment-based model can be appropriately modified 
to correctly simulate chemical systems for any 
$h \ge h_{crit}$. In particular, we also get a
measure of correctness of the original compartment-based
model.

\subsection{Molecular-based models}

\label{secDisSmol}

In this section, we study molecular-based models of reaction-diffusion 
processes, i.e. we simulate trajectories of individual molecules. 
The position $[X(t),Y(t),Z(t)]$ of a diffusing molecule (Brownian motion)
can be described by a system of three (uncoupled) stochastic 
differential equations (SDEs) \cite{Chandrasekhar:1943:SPP}
\begin{eqnarray}
X(t+\dt) 
& = &
X(t)
+
\,
\sqrt{2 D}
\;
\dW_x,
\label{xdifdW}
\\
Y(t+\dt) 
& = &
Y(t)
+
\,
\sqrt{2 D}
\;
\dW_y,
\label{ydifdW}
\\
Z(t+\dt) 
& = &
Z(t)
+
\,
\sqrt{2 D}
\;
\label{zdifdW}
\dW_z,
\end{eqnarray}
where $\dW_x$, $\dW_y$, $\dW_z$ are (uncorrelated) white
noises (i.e. differentials of the Wiener process)
and $D$ is the diffusion constant. To simulate 
trajectories of the system of SDEs 
(\ref{xdifdW})--(\ref{zdifdW}), we choose a small time
step $\Delta t$ and use the Euler-Maruyama method
to solve SDEs (\ref{xdifdW})--(\ref{zdifdW}); that is,
we compute the position $[X(t+\Delta t),Y(t+\Delta t),Z(t+\Delta t)]$
at time $t + \Delta t$ from its position $[X(t),Y(t),Z(t)]$
at time $t$ by
\begin{eqnarray}
X(t+\Delta t) 
& = &
X(t)
+
\,
\sqrt{2 D \Delta t}
\;
\xi_x,
\label{xdifxi}
\\
Y(t+\Delta t) 
& = &
Y(t)
+
\,
\sqrt{2 D \Delta t}
\;
\xi_y,
\label{ydifxi}
\\
Z(t+\Delta t) 
& = &
Z(t)
+
\,
\sqrt{2 D \Delta t}
\;
\xi_z,
\label{zdifxi}
\end{eqnarray}
where $D$ is the diffusion constant and $\xi_x$, $\xi_y$, $\xi_z$
are random numbers which are sampled from the normal distribution
with zero mean and unit variance. To model a bimolecular reaction, 
it is often postulated that two molecules (which are subject
to the bimolecular reaction) always react whenever their distance 
is less than a given reaction radius $\varrho$ 
\cite{Smoluchowski:1917:VMT,Andrews:2004:SSC}. If trajectories
of molecules exactly follow the system of SDEs 
(\ref{xdifdW})--(\ref{zdifdW}), one can find explicit formulae
linking the reaction rate constant, the diffusion constant(s)
of reactants and the reaction radius 
\cite{Smoluchowski:1917:VMT,Berg:1977:PC,Berg:1983:RWB}.
The reaction radius of heteroreaction (\ref{abexample})
is
\begin{equation}
\varrho = \frac{k_1}{4 \pi (D_A+D_B)}
\label{reactionradiussmol}
\end{equation}
and the reaction radius of homoreaction (\ref{nonlindegradationcreation})
is $k_1/(8 \pi D_A).$ Thus, the illustrative example (\ref{abexample})
can be simulated as follows. We choose a small time step 
$\Delta t$. We update the position of every molecule by 
(\ref{xdifxi})--(\ref{zdifxi}) where
$D=D_A$ for molecules of $A$ and $D=D_B$ for molecules
of $B$. Reflecting boundary 
conditions are implemented on the boundary of the cubic 
computational domain $[0,L] \times [0,L] \times [0,L]$. 
For example, if $X(t+\Delta t)$ computed by (\ref{xdifxi}) 
is less than $0$, then $X(t+\Delta t)= - X(t) - 
\sqrt{2 D \, \Delta t} \; \xi_x$. If $X(t+\Delta t)$ computed 
by (\ref{xdifxi}) is greater than $L$, then 
$X(t+\Delta t)=2 L - X(t) - \sqrt{2 D \, \Delta t} \; \xi_x$. 
Similarly for $y$ and $z$-coordinates. Whenever the distance of
a molecule of $A$ from a molecule of $B$
is less than the reaction radius
$\varrho$ given by (\ref{reactionradiussmol}), we 
remove the molecule of $A$ from the system. We also generate 
a random number $r$ uniformly distributed in $(0,1)$ during every 
time step. If $r < k_2 L^3 \Delta t$, then we generate another 
three random numbers $r_x,$ $r_y$ and $r_z$ uniformly distributed 
in $(0,1)$ and introduce a new molecule of $A$ at the position 
$(r_x L, r_y L, r_z L).$

Considering the parameter values from Figure \ref{figlargeKrdme}(a),
namely $k_1 = 0.2 \; \mu\mbox{m}^3 \; \mbox{sec}^{-1}$ and
$D_A = D_B = 1 \; \mu\mbox{m}^{2} \; \mbox{sec}^{-1}$,
and using (\ref{reactionradiussmol}),
we obtain $\varrho = 8 \, \mbox{nm}$. On the other hand, approximating
the diffusing molecule as a sphere, we can estimate its
radius by the Einstein relation \cite{Einstein:1905:UMT,Berg:1983:RWB}
\begin{equation}
\varrho_m
=
\frac{k_B T}{6 \pi \eta D},
\label{einform}
\end{equation}
where $k_B=1.38 \times 10^{-14} 
\, \mbox{g} \, \mbox{mm}^2 \, \mbox{sec}^{-2} \, K^{-1}$ 
is the Boltzmann constant, $T$ is the absolute temporature,
$\eta$ is the coefficient of viscosity and $D$ is the diffusion
constant. Considering 
a solution in water ($\eta = 10^{-3} \, \mbox{g} \;
\mbox{mm}^{-1} \, \mbox{sec}^{-1}$) at room temperature
($T=300 \, \mbox{K}$), we obtain $\varrho_m = 219.7 \, \mbox{nm}$
for $D = D_A = 1 \; \mu\mbox{m}^{2} \; \mbox{sec}^{-1}$.
Thus the reaction radius $\varrho$ given by 
(\ref{reactionradiussmol}) is unrealistically
smaller than the molecular radius $\varrho_m$. 
It is worth noting that
this undesirable property of the model does not depend
on the value of the diffusion constant $D$. 
For example, considering a hundred-times larger diffusion
constant $D$, namely 
$D_A = D_B = 100 \; \mu\mbox{m}^{2} \; \mbox{sec}^{-1}$, we obtain
the molecular radius $\varrho_m = 2.2 \, \mbox{nm}$ and 
the reaction radius $\varrho = 0.08 \, \mbox{nm}$. Let us
investigate the conditions under which the reaction radius
$\varrho$ is larger than the molecular radius $\varrho_m$.
Using (\ref{einform}), (\ref{reactionradiussmol}) 
and $D = D_A = D_B$, 
the inequality $\varrho > \varrho_m$ is equivalent to
\begin{equation*}
k_1 > \frac{4 k_B T}{3 \eta}.
\end{equation*}
Considering a solution in water at room temperature, we 
obtain that $k_1$ has to be at least of the order 
$10^8 \, \mbox{M}^{-1} \, \mbox{sec}^{-1}.$
On the other hand, typical values of $k_1$ for interactions
between proteins are of the order 
$10^6 \, \mbox{M}^{-1} \, \mbox{sec}^{-1}.$
Thus, unless the reaction rate constant $k_1$ is very 
large, the model requires the reaction radius to be
chosen unrealistically small. 
Notice that the values of the diffusion constant
$D = 1 - 100 \; \mu\mbox{m}^{2} \; \mbox{sec}^{-1}$
and the reaction rate constant
$k_1 = 0.2 \; \mu\mbox{m}^3 \; \mbox{sec}^{-1}
= 3.3 \times 10^6 \; \mbox{M}^{-1} \, \mbox{sec}^{-1}$,
which were considered previously, are in the range 
of realistic values for proteins.  

Perhaps more importantly, a
small reaction radius also provides restrictions on the
simulation time step $\Delta t$. We have to make sure that 
the average change in the distance between molecules 
during one time step, which is given by
\begin{equation}
s = \sqrt{2 (D_A + D_B) \Delta t},
\label{averchanges}
\end{equation}
is much less than the reaction radius $\varrho$,
i.e. $s \ll \varrho$ \cite{Andrews:2004:SSC}. 
Using $D_A = D_B = 10 \; \mu\mbox{m}^{2} \; 
\mbox{sec}^{-1}$ and $k_1 = 10^6 \; \mbox{M}^{-1} \, \mbox{sec}^{-1}$, 
we obtain that $\Delta t$ has to be significantly less than
a nanosecond. This limitation is even more severe for faster
diffusing molecules. Note that, in the case of the
illustrative example (\ref{abexample}), 
we also have to make sure that the production probability per 
one time step, $k_1 L^3 \Delta t$, is significantly less than
1. Considering the bimolecular reaction only, it can, in
principle, be simulated with a very large time step $\Delta t$
but the formula for $\varrho$ has to be modified accordingly.
If $s \gg \varrho$,
then the probability that a given pair of molecules interacts 
during the time step $(t,t+\Delta t)$ is proportional to the 
volume fraction
$4 \pi \varrho_\infty^3/(3 L^3)$ where $\varrho_\infty$ is
the modified reaction radius. Comparing with 
$k_1 \, \Delta t/L^3$, we obtain
\begin{equation}
\varrho_\infty 
= 
\left(\frac{3 k_1 \, \Delta t}{4 \pi} \right)^{1/3}.
\label{rhoinfinity}
\end{equation}
This formula gives a larger ($\Delta t$-dependent) reaction radius, 
but it does not have the potential to provide a spatial resolution 
close to the size of individual molecules \cite{Andrews:2004:SSC}. 
Andrews and Bray \cite{Andrews:2004:SSC} designed a computational
algorithm for intermediate values of $\Delta t$ that satisfies
$s \approx \varrho$. In this case, it is not possible
to derive an explicit formula relating $\varrho$ and $k_1$
(as was done in (\ref{reactionradiussmol}) for $s \ll \varrho$
and in (\ref{rhoinfinity}) for $s \gg \varrho$). Instead
Andrews and Bray \cite{Andrews:2004:SSC} provide
a look-up table relating (scaled) reaction
rate constant $k_1$ and reaction radius $\varrho$. However,
the reaction radius is still often smaller than molecular
radius $\varrho_m$ in their algorithm. In Section \ref{secimpsmoland},
we will modify molecular-based algorithms so that the
reaction radius can be chosen as large as the molecular 
radius. Let us note that the algorithms above consider all 
``collisions" of reactants as reactive while in reality many 
non-reactive collisions happen before the reaction takes place. 
The modified algorithms in Section \ref{secimpsmoland}
take this point into account.

\section{Improved SSAs for reaction-diffusion modelling}

\label{secImpr}

In this section, we present modified SSAs which are able
to overcome the problems mentioned in Section \ref{secDis}.
The make this section accessible to non-mathematicians,
we focus only on the results. The mathematical derivation
of the formulae presented and the justification of the modified
algorithms are given in Appendices.

\subsection{Improved compartment-based model}

\label{secImpcomp}

Let us consider the heteroreaction example (\ref{abexample})
modelled by the compartment-based reaction-diffusion model
(\ref{abexamplecompartments})--(\ref{diffusionjumpsB}). 
We will show that a suitable modification of propensity 
functions $\alpha_{ijk,1}(t)$, which were defined 
by (\ref{propfunc}), leads to an algorithm that gives the correct 
$\phi(n)$ for any $h$ larger than or equal to the critical value 
$h_{crit}$. Moreover, this is not possible for values of $h$ 
smaller than $h_{crit}.$ The critical value of the compartment
size $h$ can be estimated as 
\begin{equation}
h_{crit} = \beta_{\infty} \, \frac{k_1}{D_A + D_B}
\label{hcritform}
\end{equation}
where $k_1$ is the rate constant of the bimolecular reaction,
$D_A$ (resp. $D_B$) is the diffusion constant of $A$
(resp. $B$) and $\beta_\infty \approx 0.25272.$
If $h_{crit}$ computed by (\ref{hcritform}) is significantly
smaller than the domain size $L$, then the critical
value of $h$ is indeed given by (\ref{hcritform}). If the
domain size $L$ is comparable to (\ref{hcritform}), 
then (\ref{hcritform}) provides
only a good approximation of $h_{crit}$, with the real value being
slightly higher as discussed below. If $h \ge h_{crit}$,
we propose to modify the first formula in (\ref{propfunc})
by
\begin{equation}
\alpha_{ijk,1}(t) = A_{ijk}(t) B_{ijk}(t) \, 
\frac{(D_A+D_B) k_1}{(D_A+D_B) h^3 - \beta k_1 h^2} 
\label{propfunc2}
\end{equation}
where the parameter $\beta$ has no physical dimension and
needs to be specified. If a modeller does not have any information 
about the system, we propose to choose 
$\beta = \beta_\infty \approx 0.25272.$ We will show later
that $\beta_\infty$ is indeed the correct choice of 
$\beta$ if $K = L/h$ is large. 
Formula (\ref{propfunc2}) can be applied to any heteroreaction 
$A + B \to \emptyset$ where $\emptyset$ stands for
an arbitrary right hand side, provided that the propensity
function (\ref{propfunc2}) is positive. Consequently, the 
critical value $h_{crit}$ is the one which makes the 
denominator of (\ref{propfunc2}) equal to zero. In such
a case, the propensity function is infinity and the reaction 
happens immediately after the reacting molecules enter 
the same compartment. Thus, $h_{crit}$ satisfies
$(D_A+D_B) h_{crit}^3 - \beta h_{crit}^2 k_1 = 0$. If we substitute
$\beta_{\infty}$ for $\beta$, we obtain the
approximation (\ref{hcritform}). 

\begin{figure}
\picturesAB{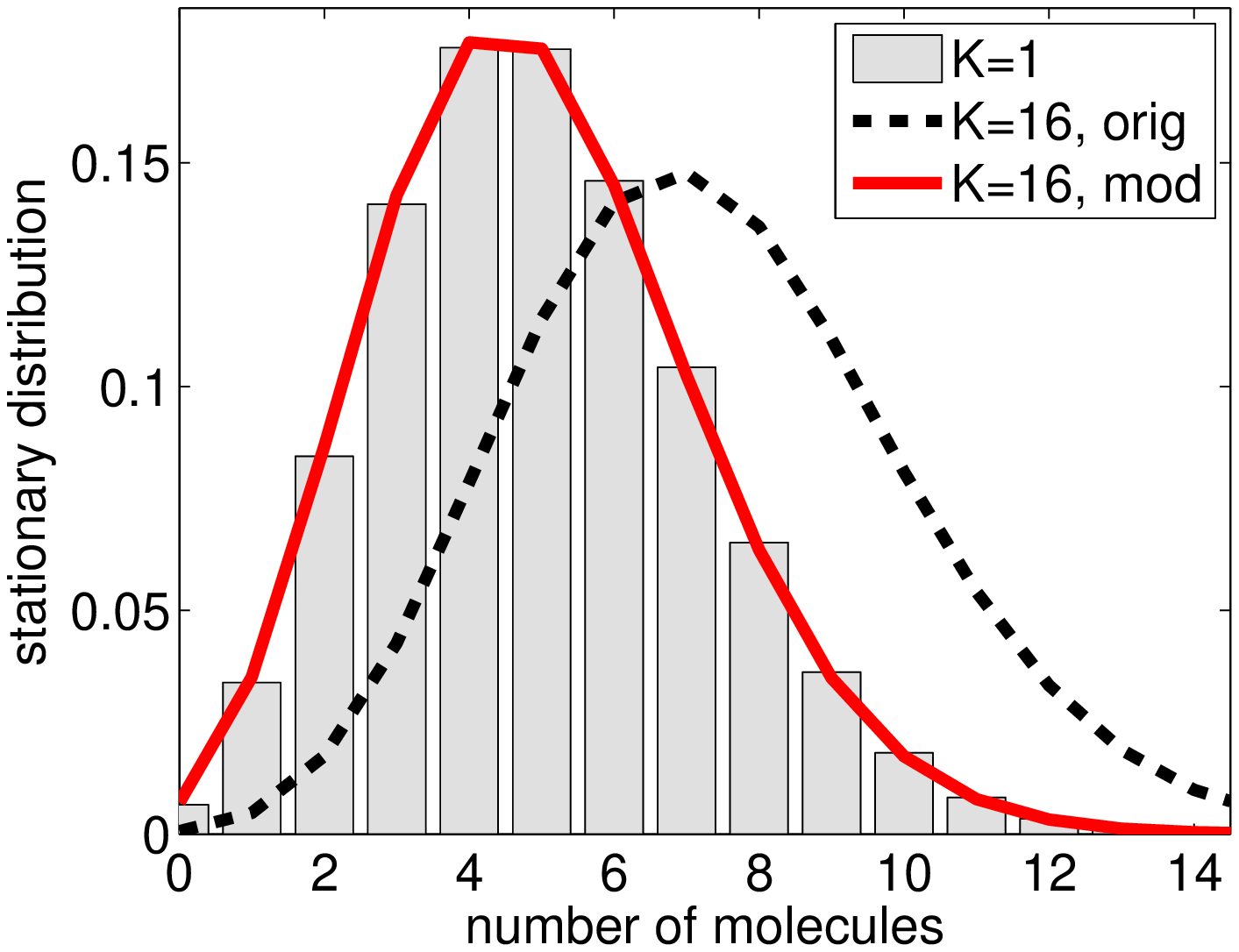}{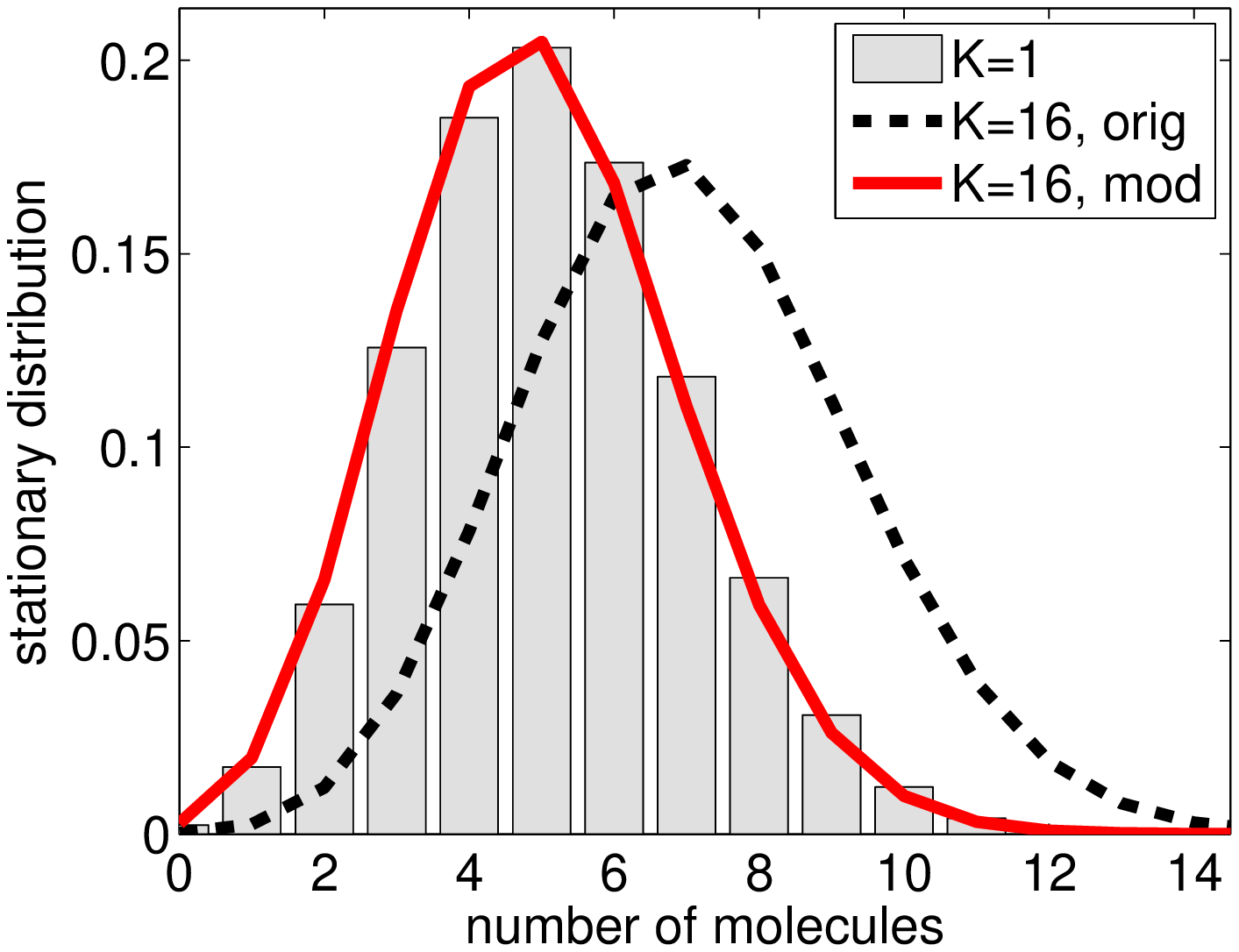}
{2.45in}{5mm}
\caption{(a) {\it Heteroreaction example $(\ref{abexample}).$
Stationary distribution $\phi_{16}(n)$ 
defined by $(\ref{statdistrpK})$ 
computed by the original SSA (dashed line)
and by the modified SSA that uses $(\ref{propfunc2})$
instead of $(\ref{propfunc})$
(solid line). Correct stationary distribution 
$\phi(n) \equiv \phi_1(n)$ is plotted as the grey histogram.}
(b) {\it Homoreaction example $(\ref{nonlindegradationcreation}).$
Stationary distribution $\phi_{16}(n)$ 
computed by the original SSA (dashed line)
and by the modified SSA that uses $(\ref{propfunc3})$
instead of $(\ref{prophomordme})$
(solid line). Correct stationary distribution 
$\phi(n) \equiv \phi_1(n)$ is plotted as the grey histogram.
}}
\label{figcorrectedRDME}
\end{figure}

Let us consider the illustrative heteroreaction example plotted in 
Figure \ref{figlargeKrdme}(a). The values of the constant 
$\beta$ for this model are given for different values of $K$ in 
Table \ref{tablebeta}, 
\begin{table}
\centerline{%
\begin{tabular}{|c|c|}
\hline
\raise -2.3mm \hbox{\rule{0pt}{7mm}}
K & $\beta$ \\
\hline 
  2 &  0.30208 \\
\hline 
  3 &  0.33233 \\
\hline 
  4 &  0.33461 \\
\hline 
  5 &  0.33119 \\
\hline 
  6 &  0.32660 \\
\hline 
  7 &  0.32205 \\
\hline 
  8 &  0.31784 \\
\hline 
\end{tabular}
\quad
\begin{tabular}{|c|c|}
\hline
\raise -2.3mm \hbox{\rule{0pt}{7mm}}
K & $\beta$ \\
\hline 
  9 &  0.31406 \\
\hline 
 10 &  0.31067 \\
\hline 
 12 &  0.30493 \\
\hline 
 14 &  0.30027 \\
\hline 
 16 &  0.29643 \\
\hline 
 18 &  0.29322 \\
\hline 
 20 &  0.29048 \\
\hline 
\end{tabular}
\quad
\begin{tabular}{|c|c|}
\hline
\raise -2.3mm \hbox{\rule{0pt}{7mm}}
K & $\beta$ \\
\hline 
 25 &  0.28514 \\
\hline 
 30 &  0.28123 \\
\hline 
 40 &  0.27587 \\
\hline 
 50 &  0.27232 \\
\hline 
 60 &  0.26979 \\
\hline 
 80 &  0.26640 \\
\hline 
100 &  0.26420 \\
\hline 
\end{tabular}
\quad
\begin{tabular}{|c|c|}
\hline
\raise -2.3mm \hbox{\rule{0pt}{7mm}}
K & $\beta$ \\
\hline 
150 &  0.26103 \\
\hline 
200 &  0.25930 \\
\hline 
300 &  0.25743 \\
\hline 
400 &  0.25643 \\
\hline 
600 &  0.25536 \\
\hline 
800 &  0.25479 \\
\hline
1000 & 0.25443 \\
\hline 
\end{tabular}
}
\caption{{\it The values of $\beta$ for the selected values of $K$ 
computed by $(\ref{betaformula})$ for the heteroreaction example 
$(\ref{abexample})$ for $B_0 = 1$.}}
\label{tablebeta}
\end{table}
and lie between $\beta_\infty \approx 0.25272$ and $0.34$. 
Increasing $K$ to infinity, the values of $\beta$ converge 
to $\beta_\infty$. In Figure \ref{figcorrectedRDME}(a), 
we compare the results computed
using the original formula (\ref{propfunc}) and by the new 
formula (\ref{propfunc2}) for the heteroreaction example
(\ref{abexample}). We use the same parameter values as in
Figure \ref{figlargeKrdme}(a) and $K=16$.
As in Figure \ref{figlargeKrdme}(a), we observe a difference 
between $\phi_1(n)$ and $\phi_{16}(n)$ if the original model
is used. On the other hand, $\phi_{16}(n)$ computed by
the modified algorithm (solid line) is the same as 
$\phi_1(n)$ (grey histogram). 

In Table \ref{tablebeta}, we observe that $\beta$ weakly
depends on $K = L/h$, so that the real value of $h_{crit}$
(that makes the propensity function (\ref{propfunc2}) 
equal to infinity) is slightly larger than (\ref{hcritform}).
The dependence of $\beta$ on $K$ is caused by the boundary 
of the computational domain $[0,L] \times [0,L] \times [0,L]$.
Every inner compartment can be entered from six possible
directions, but some incoming directions are missing
in the boundary compartments. Whenever a molecule of $B$ is in a
boundary compartment, it is less likely found by molecules
of $A$. One could address this problem either
by introducing different propensity functions for different 
compartments, or by using $\beta$ in (\ref{propfunc2})
which is slightly larger than $\beta_{\infty}$. We used the
latter option. In \ref{appenC}, we show that the modified
$\beta$ is given by
\begin{equation}
\fl
\beta
=
\frac{1}{2K^3} \!\!\!
\sum_{\hbox{\hsize=0.145\hsize\vbox{
      \noindent
      \scriptsize
      \;\;\;\;\;$i,j,k=0$ \\
      $(i,j,k) \ne (0,0,0)$}}}^{K-1} 
\frac{1}{
3
-
\cos \left( i \pi/K \right)
-
\cos \left( j \pi/K \right)
-
\cos \left( k \pi/K \right)
}.
\label{betaformula}
\end{equation}
The results in Table \ref{tablebeta} have been computed
by (\ref{betaformula}). In \ref{appdercomp}, we 
show that $\beta_\infty$ is given by
\begin{equation}
\beta_\infty 
=
\frac{1}{2 \pi^2}
\int_0^{\pi}
\int_0^{\pi}
\frac{1}{\sqrt{(3 - \cos x - \cos y)^2 - 1}} \,
\mbox{d}x \,
\mbox{d}y.
\label{betainfinityformula}
\end{equation}
Evaluating this integral numerically by the Monte Carlo method,
we obtain $\beta_\infty \approx 0.25272$. If we model a complicated
reaction-diffusion system, modelling of each heteroreaction 
will be improved by using (\ref{propfunc2}), provided
that all rates obtained by (\ref{propfunc2}) are positive. 
In other words, the smallest possible $h$ which can be 
simulated is given as the maximal $h_{crit}$ for each
bimolecular reaction. 
If $h$ is significantly larger then $h_{crit}$ (i.e. 
if $h \gg h_{crit}$), then we have $(D_A + D_B) h^3 \gg \beta k_1 h^2$ 
and we can approximate
\begin{equation}
\frac{(D_A+D_B) k_1}{(D_A+D_B) h^3 - \beta k_1 h^2}
\approx
\frac{k_1}{h^3}.
\label{largehk1overh3}
\end{equation}
In particular, the propensity function $\alpha_{ijk,1}(t)$
defined by (\ref{propfunc2}) is approximately equal 
to the original propensity function (\ref{propfunc})
for large values of $h$. On the other hand, if $h$ is close
to $h_{crit}$, then the propensity function
$\alpha_{ijk,1}(t)$ given by (\ref{propfunc2}) 
is larger than the original propensity function (\ref{propfunc}).

Finally, let us consider the homoreaction example
(\ref{nonlindegradationcreation}) modelled by the 
compartment-based model. In this case, we propose
to modify the propensity functions (\ref{prophomordme})
for $h \ge h_{crit}$, by
$\alpha_{ijk,1}(t)$
\begin{equation}
\alpha_{ijk,1}(t) = A_{ijk}(t) (A_{ijk}(t)-1) \, 
\frac{D_A k_1}{D_A h^3 - \beta k_1 h^2}, 
\label{propfunc3}
\end{equation}
where $\beta$ is a constant. In Figure \ref{figcorrectedRDME}(b), 
we compare the results computed using the original formula 
(\ref{prophomordme}) and by the new formula (\ref{propfunc3}) 
for the value of $\beta$ given by Table \ref{tablebeta}. We 
use the same parameter values as in Figure \ref{figlargeKrdme}(b) 
and $K=16$. We again observe that the modified formula 
(\ref{propfunc3}) gives better results than the original
SSA. One can still observe a small error which is caused by the fact 
that we used the value of $\beta$ computed for heteroreactions. 
Using (\ref{betaformula}), we have $\beta \approx 0.29643$ for $K=16$ 
(see Table \ref{tablebeta}). Experimenting with the model
(\ref{nonlindegradationcreation}), we can find that $\beta = 0.28$ 
yields slightly better fit between $\phi_1(n)$ and $\phi_{16}(n)$.
Notice that $\beta = 0.28$ is still larger than $\beta_\infty \approx 0.25272$.
However, for the purposes of applications, it is sufficient
to use either $\beta = \beta_\infty$ or the values of $\beta$ from 
Table \ref{tablebeta} for both heteroreactions and homoreactions. 
Using $\beta = \beta_\infty$, we discovered that the biggest contribution 
of the error stems from the boundary effects and derived Table 
\ref{tablebeta} which adds a correction to $\beta_\infty$ to compensate 
for boundary behaviour. In a similar way, one could look for further 
corrections to the value of $\beta$ for homoreactions, or for domains 
which are cuboids 
rather than cubes. Although such corrections are of interest from the 
mathematical point of view, they provide only a negligible improvement 
of the algorithm. Thus we will not include them in this paper.

\subsection{Improved molecular-based models}

\label{secimpsmoland}

The major assumption of molecular-based models is that molecules always 
react whenever their distance is less than the reaction radius 
$\varrho$. The reaction radius $\varrho$ is related to the rate constant
of the bimolecular reaction by a simple formula (for example, 
(\ref{reactionradiussmol}) for the Smoluchowski model) or by 
a look up table for the Andrews and Bray model \cite{Andrews:2004:SSC}.
In this section, we present models that implement bimolecular
reactions with the help of two parameters: the reaction radius
$\newrho$ and the reaction rate $\lambda$. We postulate
that the bimolecular reaction can take place only when the distance of
molecules is less than $\newrho$. If this is the case,
then the bimolecular reaction events happen with the rate $\lambda$. 
We will call this model $\lambda-\newrho$ model in what follows.

To implement this idea on the computer, we need to relate 
the parameters $\lambda$ and $\newrho$ to the rate constant
of the bimolecular reaction. The advantage of the $\lambda-\newrho$ 
model is that many different pairs of $\lambda$ and $\newrho$
correspond to the same bimolecular rate constant. In mathematical
terms, the condition on  $\lambda$ and $\newrho$ is one equation for 
two unknowns $\lambda$ and $\newrho$. In particular, we can choose
the value of $\newrho$ as desired (e.g. to be comparable to the molecular 
radius $\varrho_m$) and compute the appropriate value of $\lambda$. Thus
$\lambda-\newrho$ model has the potential to solve the problems of
molecular-based modelling discussed in Section \ref{secDisSmol}.

We explain the $\lambda-\newrho$ model on the heteroreaction example 
(\ref{abexample}). The diffusion of molecules $A$ and $B$ is
simulated as in Section \ref{secDisSmol}. We choose a small time 
step $\Delta t$. The trajectory of every molecule is computed
by (\ref{xdifxi})--(\ref{zdifxi}) where $D=D_A$ for molecules of 
$A$ and $D=D_B$ for molecules of $B$. 
Let $s$ be the average change in the relative position of a molecule of
$A$ and a molecule of $B$ during one time step, given
by (\ref{averchanges}).
We will distinguish two cases of the value of the time step:
(i) the time step $\Delta t$ is chosen so small that $s \ll \newrho$;
and (ii) the time step $\Delta t$ is larger so that $s \approx \newrho$.

\medskip

\noindent
{\it (i) Small time step $\Delta t$.} To model heteroreaction $A + B \to
\cdots$, we introduce two parameters: reaction radius $\newrho$ and rate 
$\lambda$. The reaction radius is expressed in units of length and rate 
$\lambda$ in units per time. Whenever the distance of a molecule of $A$ 
from a molecule of $B$ is less than the reaction radius $\newrho$, 
then the heteroreaction  takes place with the rate $\lambda$. 
In \ref{appderk1lambda}, we derive the following relation between $\newrho$, 
$\lambda$ and the rate constant $k_1$ of the heteroreaction:
\begin{equation}
k_1
=
4 \pi (D_A+D_B)
\left(
\newrho
-
\sqrt{\frac{D_A+D_B}{\lambda}} \, 
\tanh \left( \newrho \, \sqrt{\frac{\lambda}{D_A+D_B}} \right)
\right).
\label{formk1lambda}
\end{equation}
This is one condition for two unknowns $\newrho$ and $\lambda$.
In particular, we can choose $\newrho$ comparable to the radii of
reacting molecules and use (\ref{formk1lambda}) to compute the 
corresponding $\lambda$. Notice that (\ref{formk1lambda})
is a simple non-linear equation which can be solved by any
numerical method for finding roots of a real-valued function
(for example, Newton's method or the bisection method). 

If $\lambda = \infty$ (that is, if molecules
react immediately whenever they are within the reaction 
radius), then (\ref{formk1lambda}) simplifies
to (\ref{reactionradiussmol}) as desired. On the other
hand, if $\lambda$ is small that
$\lambda \ll(D_A+D_B) \, \newrho^2$, then we can use
Taylor expansion in (\ref{formk1lambda}) to approximate
\begin{equation*}
\tanh \left( \newrho \, \sqrt{\lambda/(D_A+D_B)} \right)
\approx
\newrho \, \sqrt{\lambda/(D_A+D_B)} 
-
\frac{1}{3} \left( \newrho \, \sqrt{\lambda/(D_A+D_B)} \right)^3
\!.
\end{equation*}
Consequently, (\ref{formk1lambda}) simplifies to
$k_1 \approx 4 \pi \newrho^3 \lambda/3$ which can be 
equivalently rewritten as 
\begin{equation}
\lambda \approx \frac{k_1}{4 \pi \newrho^3 /3},
\label{lambdasmall}
\end{equation}
i.e. the reaction rate $\lambda$ is given as the reaction rate constant
$k_1$ divided by the volume, $4 \pi \newrho^3 /3$, of
the ball in which the reaction takes place. Formula (\ref{lambdasmall})
is analogous to the formula for the reaction rate per compartment
in the compartment-based approach for large compartment size
$h$. If $h$ is large satisfying $h \gg h_{crit}$, then the reaction rate 
per compartment is given as $k_1/h^3$, which is the  reaction rate 
constant $k_1$ divided by the volume, $h^3$, of the compartment
-- see (\ref{largehk1overh3}).

\medskip

\noindent
{\it (ii) SSA for larger time steps.}
We introduce two parameters: reaction radius $\newrho$ and
probability $P_\lambda$. The heteroreaction $A + B \to \cdots$ 
is modelled as follows: whenever the distance between a molecule 
of $A$ and a molecule of $B$ (at the end of a time step)
is less than the reaction radius $\newrho$, then the heteroreaction 
takes place with probability $P_\lambda$; that is, we generate 
a random number $r$ uniformly distributed in $(0,1)$ and the 
heteroreaction (removal/addition of molecules) is performed 
whenever $r < P_\lambda$. Notice that the previous algorithm
(for small time step $\Delta t$) can be also formulated in 
terms of parameters $\newrho$ and $P_\lambda$ rather than 
$\newrho$ and $\lambda$. Indeed, if $\lambda \, \Delta t \ll 1$, 
we have $P_\lambda \approx \lambda \, \Delta t$. If $\Delta t$
is larger, then the relation between $P_\lambda$ and $\lambda$
is more complicated. However, from the practical point of view,
there is no need to know the rate $\lambda$: it is sufficient
to formulate the algorithms in terms of $\newrho$ and $P_\lambda$.
Next, we present the condition relating $\newrho$ and $P_\lambda$
with the rate constant $k_1$ of heteroreaction $A + B \to \cdots$.
We define the dimensionless parameters 
\begin{equation}
\gamma
=
\frac{s}{\newrho}
=
\frac{\sqrt{2 (D_A + D_B) \Delta t}}{\newrho},
\qquad\qquad
\kappa
=
\frac{k_1 \, \Delta t}{\newrho^3}.
\label{gammakappaparameter}
\end{equation} 
In applications, we first specify the time step $\Delta t$. We also 
want to specify $\newrho$ in a realistic parameter range. 
Consequently, $\gamma$ and $\kappa$ can be considered as given numbers
in what follows. For example, we can choose the average change of
distance between reacting molecules during one time step equal
to the reaction radius, i.e. $s = \newrho$. Then 
(\ref{gammakappaparameter}) gives $\gamma = 1$.
The key modelling question is: what is the appropriate
value of the probability $P_\lambda$?
\begin{figure}
\picturesAB{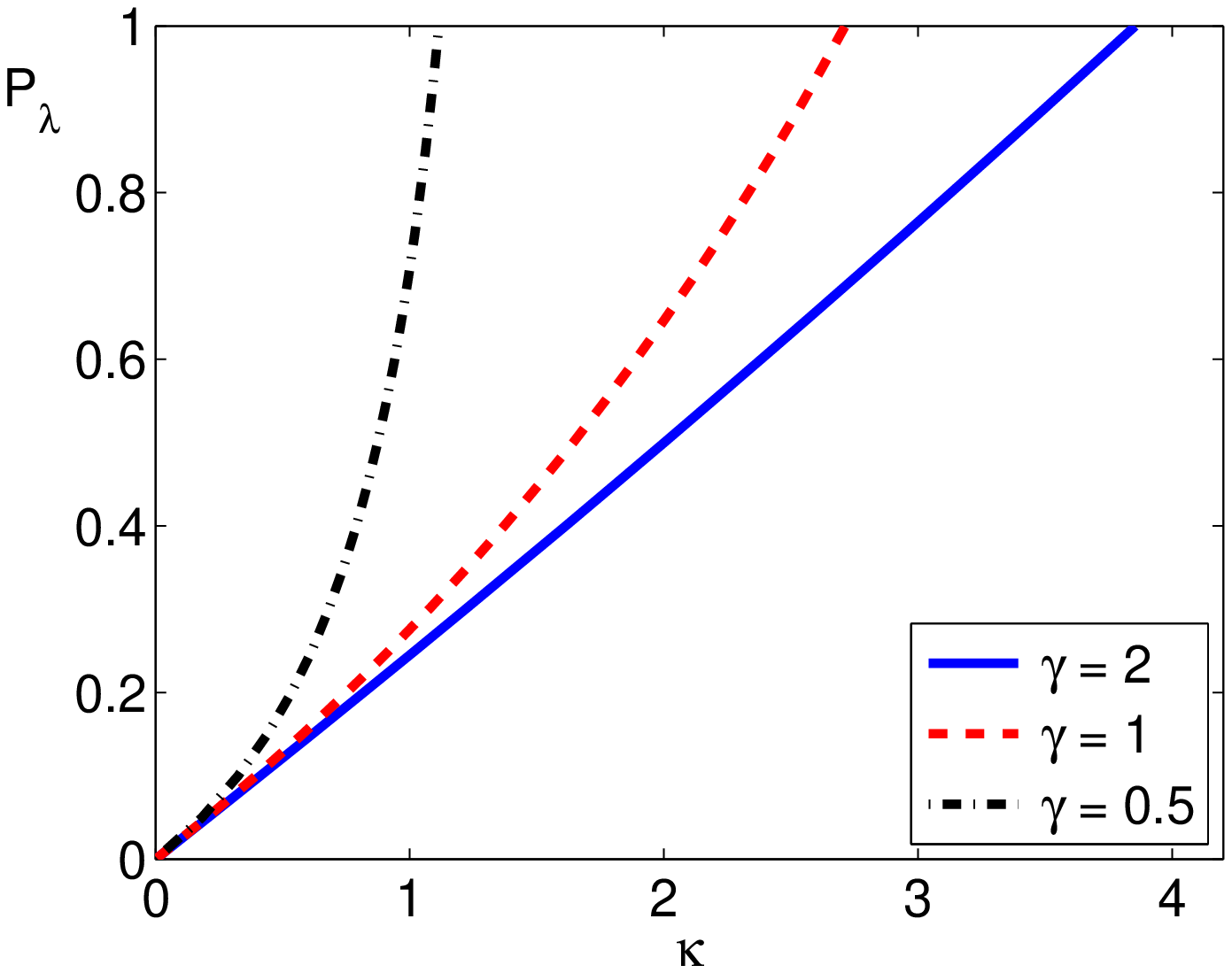}{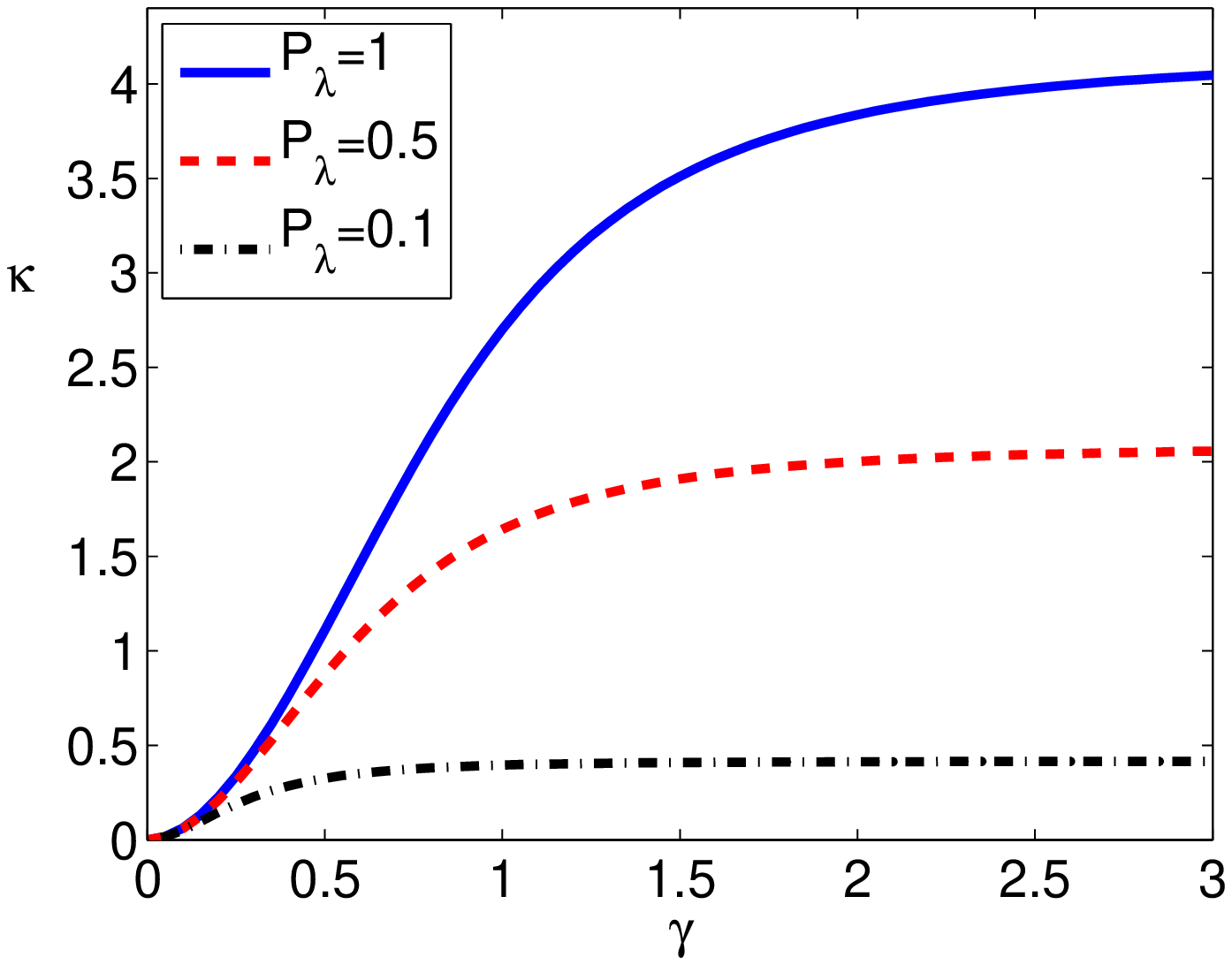}
{2.45in}{5mm}
\caption{(a) {\it Dependence of $P_\lambda$ on
$\kappa$ for three different values of $\gamma$.
Dimensionless parameters $\kappa$ and $\gamma$ 
are given by $(\ref{gammakappaparameter})$.}
(b) {\it Dependence of $\kappa$ on $\gamma$
for three different values of $P_\lambda$.
}}
\label{figdependencekPs}
\end{figure}
In Figure \ref{figdependencekPs}(a), we present the dependence of
$P_\lambda$ on $\kappa$ for three different values
of $\gamma$. The derivation of the equation for $P_\lambda$ and
the numerical method which was used to compute this
plot are given in \ref{appendF}. Below, we summarize only the
equations that were solved and present illustrative computational results.

To formulate the equation for $P_\lambda$, it is useful to define
an (auxiliary) function $g(r): [0,\infty) \to [0,1]$ 
as the solution of the integral equation
\begin{equation}
g(r) 
=
(1-P_\lambda)
\int_0^1 
K(r,r^\prime; \gamma) \,
g(r^\prime)
\, \dr^\prime
+
\int_1^\infty 
K(r,r^\prime; \gamma) \,
g(r^\prime)
\, \dr^\prime,
\label{IntegrEquation2}
\end{equation}
satisfying $g(r) \to 1$ as $r \to \infty$, where
\begin{equation}
K(r,r^\prime; \gamma)
=
\frac{r^\prime}{r \gamma \sqrt{2 \pi}}
\left(
\exp \left[ - \frac{(r-r^\prime)^2}{2 \gamma^2} \right]
-
\exp \left[ - \frac{(r+r^\prime)^2}{2 \gamma^2} \right]
\right).
\label{defKrr}
\end{equation}
The function $g(r)$ depends on dimensionless parameters 
$P_\lambda$ and $\gamma$; we make this explicit by writing 
$$
g(r; P_\lambda, \gamma) \equiv g(r).
$$ 
Then, the model parameters $\newrho$,
$P_\lambda$, $\Delta t$ are related to rate constant
$k_1$ and diffusion constants $D_A$, $D_B$ by  
\begin{equation}
\kappa
=
P_\lambda
\int_0^1 
4 \pi r^2
g(r; P_\lambda, \gamma)
\, \dr.
\label{imporequation}
\end{equation}
Since $k_1$, $D_A$ and $D_B$ are known and parameters
$\Delta t$ and $\newrho$ can be specified first, 
parameters $\gamma$ and $\kappa$ are in applications 
given numbers.
Thus (\ref{imporequation}) is one equation for one unknown 
$P_\lambda$. In \ref{appendF}, we present a numerical
approach for solving this equation, as well as the derivation
of (\ref{IntegrEquation2})--(\ref{imporequation}).

In Figure \ref{figdependencekPs}(b), we present the dependence of
$\kappa$ on $\gamma$ for three different values of $P_\lambda$. 
Note that the case $P_\lambda = 1$ corresponds to the Andrews
and Bray model \cite{Andrews:2004:SSC}. Thus the solid line 
in Figure \ref{figdependencekPs}(b) has been already computed 
in reference \cite{Andrews:2004:SSC}. However, we propose
to use a much smaller value of  $P_\lambda$, which enables 
us to choose a larger (more physically meaningful) reaction radius. 
We see in Figure \ref{figdependencekPs} that reducing $P_\lambda$
at $\gamma$ fixed corresponds to reducing $\kappa$, thereby
increasing the reaction radius. 

\begin{figure}
\picturesAB{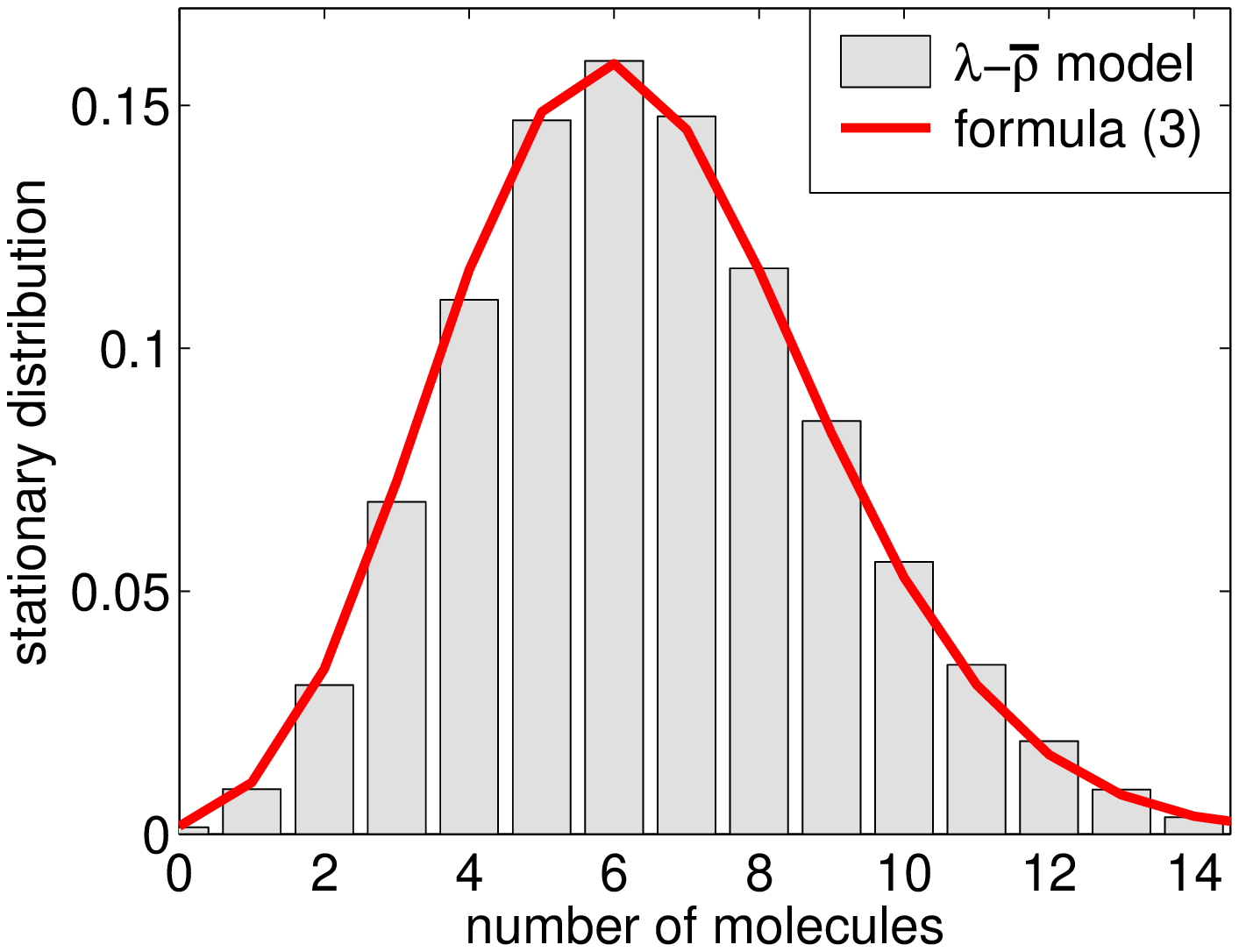}{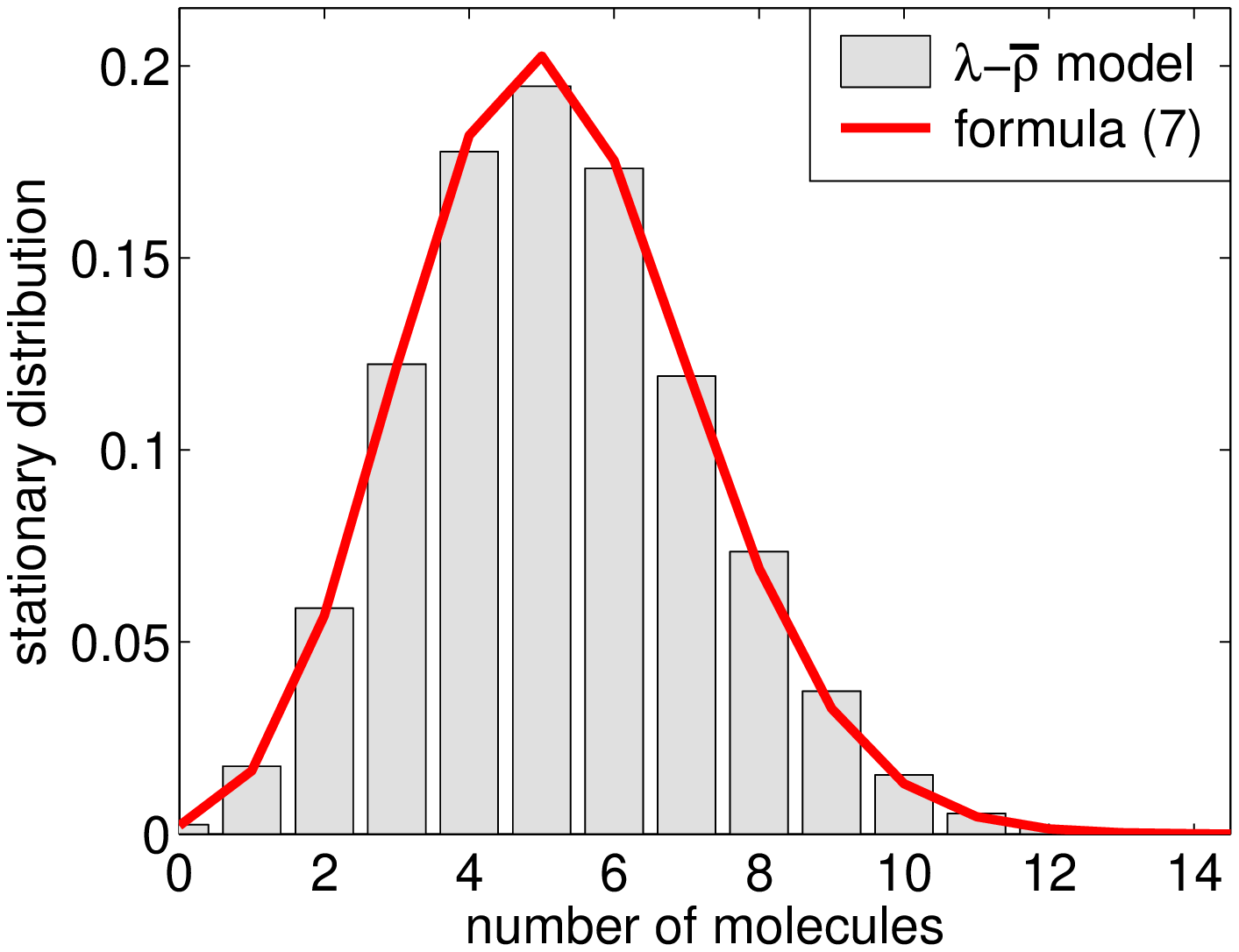}
{2.45in}{5mm}
\caption{(a) {\it Heteroreaction example $(\ref{abexample}).$
Stationary distribution computed by
the $\lambda$-$\newrho$ model (grey histogram)
and by formula $(\ref{phindistrabexample})$ (solid line).
We use 
$k_1 = 0.2 \; \mu\mbox{m}^3 \; \mbox{sec}^{-1}$,
$k_2 = 0.02 \; \mu\mbox{m}^{-3} \; \mbox{sec}^{-1}$,
$D_A = D_B = 10 \; \mu\mbox{m}^{2} \; \mbox{sec}^{-1}$,
$L = 2 \; \mu\mbox{m}$ and $B_0=1$, $\Delta t = 10^{-5} \; \mbox{sec}$,
$\newrho = 40 \; \mbox{nm}$ and $P_\lambda = 7.7 \times 10^{-3}$.}
(b) {\it Homoreaction example $(\ref{nonlindegradationcreation}).$
Stationary distribution computed by the $\lambda$-$\newrho$ model 
(grey histogram) and by formula $(\ref{pnmgfstatGs})$ (solid line). 
We use 
$k_1 = 0.1 \; \mu\mbox{m}^3 \; \mbox{sec}^{-1}$,
$k_2 = 0.08 \; \mu\mbox{m}^{-3} \; \mbox{sec}^{-1}$,
$D_A = 10 \; \mu\mbox{m}^{2} \; \mbox{sec}^{-1}$
and $L = 2 \; \mu\mbox{m}$,
$\newrho = 40 \; \mbox{nm}$ and 
$P_\lambda = 7.7 \times 10^{-3}$.
}}
\label{figSDEexample}
\end{figure}

The heteroreaction example (\ref{abexample}) is simulated
by the $\lambda-\newrho$ model as follows. We update the position
of every molecule by (\ref{xdifxi})--(\ref{zdifxi}) where $D=D_A$ for 
molecules of $A$ and $D=D_B$ for molecules of $B$. Reflecting boundary 
conditions are implemented on the boundary of the cubic computational domain 
$[0,L] \times [0,L] \times [0,L]$ as in Section \ref{secDisSmol}.
The production of molecules of $A$ (i.e. the second reaction in 
(\ref{abexample})) is simulated as before. We generate a random number 
$r$ uniformly distributed on $[0,1]$ during every time step. 
If $r < k_2 L^3 \Delta t$, 
then we generate another three random numbers $r_x,$ $r_y$ and $r_z$ 
uniformly distributed on $[0,1]$ and introduce a new molecule of $A$ 
at the position $(r_x L, r_y L, r_z L).$ In particular $\Delta t$,
has to be chosen small enough that $k_2 L^3 \Delta t \ll 1.$
If the separation between a molecule of $A$ and a molecule of $B$ 
(at the end of a time step)
is less than the reaction radius $\newrho$, then we generate 
a random number $r$ uniformly distributed on $[0,1]$ and we
remove the molecule of $A$ from the system if $r < P_\lambda$. 
In Figure \ref{figSDEexample}(a), we present
the stationary distribution computed by this algorithm (grey histogram). 
The value of $P_\lambda$, i.e $P_\lambda = 0.77 \, \%$, was computed
by solving (\ref{IntegrEquation2})--(\ref{imporequation}) using 
the numerical method given in \ref{appendF}. The values of parameters
are given in the caption of Figure \ref{figSDEexample}(a). Notice
that the reaction radius is $40$ nm and the time step $\Delta t$ was 
chosen so that $\gamma = 0.5$. The comparison of computational results
with formula (\ref{phindistrabexample}) (solid line) is excellent.

Finally, let us discuss the homoreaction example 
$(\ref{nonlindegradationcreation}).$ Since two molecules of
$A$ are removed from the system whenever the homoreaction takes
place, we have to replace $k_1$ by $2 k_1$ in the above
formulae. In particular, we replace $\kappa$ by $2 \kappa$
in (\ref{imporequation}). Moreover, $D_A + D_B$ has to be 
replaced by $2 D_A$ in all formulae. Otherwise, method
(\ref{IntegrEquation2})--(\ref{imporequation}) 
for computing $P_\lambda$ stays the same.  
In Figure \ref{figSDEexample}(a), we present
the stationary distribution computed by the $\lambda$-$\newrho$ model 
(grey histogram). The comparison with exact formula $(\ref{pnmgfstatGs})$ 
(solid line) is again excellent. 

\section{Discussion}

\label{secdiscussion}

In this paper, we used the illustrative examples (\ref{abexample})
and (\ref{nonlindegradationcreation}) to compare the
results of different stochastic reaction-diffusion methods.  
The advantage of illustrative chemical 
models (\ref{abexample}) and (\ref{nonlindegradationcreation})
is that they have non-trivial stationary distributions 
given by (\ref{phindistrabexample}) and (\ref{pnmgfstatGs}),
respectively. In principle, one could study $A + B \to B$ 
(or $A + A \to B$) on its own to make the illustrative examples 
even simpler. However, the number of molecules of $A$ would then 
decrease to zero as time progresses and the stationary distribution 
would be trivial (i.e. there would be 0 molecules with probability 
1 in the system after long time). The trivial stationary distribution 
is obtained for $A + B \to B$ (or $A + A \to B$) by any 
reaction-diffusion SSA, so we
would not learn anything useful from the stationary behaviour.
We would observe differences in modelling the transient behaviour of 
bimolecular reactions. However, the transient
behaviour depends on the initial condition. For these reasons, 
we coupled bimolecular reactions with the production of the chemical 
species $A$ to obtain the model chemical systems 
(\ref{abexample}) and (\ref{nonlindegradationcreation}) which
have the non-trivial stationary distributions. 
It is worth noting that the model systems in this paper do 
not have any spatial variation of the probability distribution. 
No part of the computational domain is preferred by molecules of 
$A$ or $B$ and the resulting probability distribution is homogeneous 
in space. It is easy to generalize (\ref{abexample}) and 
(\ref{nonlindegradationcreation})
to the spatially non-homogeneous case (for example, by considering 
production reaction $\emptyset \to A$ only in part of the 
computational domain
\cite{Erban:2007:PGS}). Such a generalization is necessary
for studying some other aspects of reaction-diffusion SSAs 
which we will address in a future publication.
However, our examples (\ref{abexample}) and 
(\ref{nonlindegradationcreation}) were complex enough to illustrate 
all results of this paper. 

We studied both on-lattice and off-lattice SSAs for reaction-diffusion 
processes. In particular, we were able to see connections
between both types of models. For on-lattice models we found that
there was a limitation on the compartment
size $h$ from below, i.e. $h \ge h_{crit}$. 
In Section \ref{secImpcomp}, we showed that the rate of bimolecular
reaction per compartment must be chosen to be infinity for $h = h_{crit}$.
In a similar way for the off-lattice model, 
a decrease of $\newrho$ in the $\lambda$-$\newrho$ 
model, presented in Section \ref{secDisSmol}, must be compensated
by an increase in the rate $\lambda$ (probability $P_\lambda$).
Again, there is a limitation on $\newrho$ from below,
i.e. $\newrho$ must be larger than or equal to the $\varrho$ of
the Smoluchowski model which is given by (\ref{reactionradiussmol}). 
If $\newrho$ is sufficiently larger than this limiting
value, then $\lambda$ is given by (\ref{lambdasmall}),
that is, the reaction rate constant $k_1$ divided by the volume, 
$4 \pi \newrho^3 /3$, of the ball in which the reaction takes place. 
This is analogous to the situation $h \gg h_{crit}$ where
the reaction rate per compartment is given by $k_1/h^3$
(reaction rate constant $k_1$ divided by the volume, $h^3$, of the 
compartment). Thus both models give, in the limit of large
$\newrho$ and large $h$, the same expression for local
reaction rates: rate constant divided by the volume in which
the reaction takes place. 

The results of this paper have been summarized in Section \ref{secImpr}.
They were explained on illustrative computational examples, but
the general formulae can be applied to modelling
bimolecular reactions which are part of complex reaction-diffusion 
processes. We presented our results as improvements of two commonly 
used reaction-diffusion SSAs which have been previously implemented
in reaction-diffusion software packages MesoRD \cite{Hattne:2005:SRD} 
and Smoldyn \cite{Andrews:2004:SSC}. Other molecular-based
models, such as MCell \cite{Stiles:2001:MCM}, 
Green's-function reaction dynamics \cite{vanZon:2005:GFR}
and velocity jump processes \cite{Erban:2007:RBC,Erban:2004:ICB},
were not directly studied in this paper but some of the
ideas presented in Section \ref{secimpsmoland} can be 
applied to improve them too. From the application point
of view, we focussed on modelling bimolecular reactions 
of biomolecules, e.g. proteins, but the concepts presented can be 
also applied to stochastic reaction-diffusion modelling in 
population ecology \cite{McKane:2004:SMP} or to modelling
cellular dispersal \cite{Erban:2005:STS,Erban:2007:TEA}.
In these cases, the diffusing objects are not macromolecules but
cells or animals, and the bimolecular ``reaction" is not 
a chemical reaction but local interaction between two cells
or animals, for example, competition or predation \cite{McKane:2004:SMP}. 

\section*{Glossary}

\noindent
{\it Gillespie SSA.} Stochastic simulation algorithm for simulating
the time-evolution of well-stirred chemical systems. The results
are consistent with the solution of the chemical master equation
\cite{Gillespie:1977:ESS,Gillespie:1992:MPI}.  

\medskip

\noindent
{\it Markov Chain.} Stochastic process for which the future states
of the system only depend on the present state and are independent
of the past states.

\medskip

\noindent
{\it Modified Bessel function of the first kind.} The evaluation of modified 
Bessel functions is part of any standard mathematical software (e.g. the 
function {\tt besseli} in Matlab). In general, the modified Bessel 
function $I_n$ (for $n \in {\mathbb N}$) is a solution of the 
ordinary differential equation
\begin{equation*}
z^2 \, I_n^{\prime\prime}(z) + z \, I_n^{\prime}(z) - (z^2 + n^2) \, I_n(z)
=
0.
\end{equation*}

\section*{Acknowledgments}

This publication is based on work supported by Award No. KUK-C1-013-04 , 
made by King Abdullah University of Science and Technology (KAUST).
This work was also partially supported by Somerville College, 
University of Oxford.

\appendix

\section{Stationary distributions, means and variances for
the illustrative heteroreaction and homoreaction examples}
	 
\label{secmathanalmod}

Let us consider chemical system (\ref{abexample}) to be well-stirred. 
Let $p_n(t)$ be the probability that there are $n$ molecules of $A$ 
at time $t$ in the reactor, i.e. $A(t)=n$. Then $p_n(t)$ evolves 
according to the 
chemical master equation \cite{Erban:2007:PGS,Gillespie:2000:CLE}
\begin{equation}
\frac{\mbox{d} p_n}{\dt}
=
\frac{k_1 B_0}{\vol} 
\, (n+1) \, p_{n+1} - 
\frac{k_1 B_0}{\vol} 
\, n \, p_n
+
k_2 \vol \, p_{n-1} - k_2 \vol \, p_n 
\label{cmeabexample}
\end{equation}
where the third term on the right hand side is missing in 
(\ref{cmeabexample}) for $n=0$; i.e. we use the convention that
$p_{-1} \equiv 0.$ The stationary distribution $\phi(n)$ 
is defined by 
\begin{equation}
\phi(n) = \lim_{t \to \infty} p_n(t).
\label{statdistrp}
\end{equation}
Consequently, (\ref{cmeabexample}) implies that $\phi(n)$ 
satisfies the equation
\begin{equation*}
\frac{k_1 B_0}{\vol} 
\, (n+1) \, \phi(n+1) - 
\frac{k_1 B_0}{\vol} 
n \, \phi(n)
+
k_2 \vol \, \phi(n-1) 
- 
k_2 \vol \, \phi(n)
=
0 
\end{equation*}
where $\phi(-1)=0$, which can be equivalently written as
\begin{eqnarray}
\phi(1)
&=&
\frac{k_2 \vol^2}{k_1 B_0} \, \phi(0), 
\nonumber
\\
\phi(n)
&=&
\left(
\frac{k_2 \vol^2}{k_1 B_0 n} 
+ 1 - \frac{1}{n}  
\right)
\, \phi(n-1)
-
\frac{k_2 \vol^2}{k_1 B_0 n} \, \phi(n-2),
\;\; \mbox{for} \; n \ge 2.
\label{pomcmephi}
\end{eqnarray}
Thus $\phi(n)$ is uniquely determined by the value 
of $\phi(0)$. We can easily verify 
that (\ref{phindistrabexample}) satisfies 
(\ref{pomcmephi}). Moreover, it is the only solution of 
(\ref{pomcmephi}) that satisfies the normalization 
condition $\sum_{n=0}^{\infty} \phi(n) = 1$.
The stationary value of the stochastic mean $M_s$ (i.e. the value 
around which the number of molecules fluctuates) and the stationary 
value of the variance $V_s$ (i.e. the size of the stochastic 
fluctuations) are given by
\begin{equation}
M_s
=
\sum_{n=0}^{\infty} n \, \phi(n),
\qquad
V_s
=
\sum_{n=0}^{\infty} \big( n - M_s \big)^2 \, \phi(n).
\label{defMsVs}
\end{equation}
Using (\ref{phindistrabexample}), we obtain (\ref{meanMsabexample})
and $V_s = M_s$.

Let us consider the homoreaction example (\ref{nonlindegradationcreation}).
Then the chemical master equation reads as follows
\begin{equation*}
\frac{\mbox{d} p_n}{\dt}
=
\frac{k_1}{\vol} 
\, (n+2) (n+1) \, p_{n+2} - 
\frac{k_1}{\vol} 
\, n (n-1) \, p_n
+
k_2 \vol \, p_{n-1} - k_2 \vol \, p_n. 
\end{equation*}
Starting with the stationary version of this equation,
one can use the method of moment generating function 
\cite{vanKampen:2007:SPP} to show that the stationary 
distribution $\phi(n)$ is given by
(\ref{pnmgfstatGs}) \cite{Engblom:2006:CMH,Twomey:2007:SMR}.
The stationary values of stochastic mean $M_s$ and variance $V_s$, 
which are defined by (\ref{defMsVs}),
can be also evaluated in terms of the Bessel functions;
$M_s$ is given by (\ref{formulaMsGs}) and 
$V_s = M_s - M_s^2 + k_2 \vol^2/(2 k_1).$ The classical 
deterministic description of the chemical system 
(\ref{nonlindegradationcreation}) is given, for concentration 
$a(t)=A(t)/\vol$, as the ODE
$
\mbox{d}a/\dt
=
- 2 k_1 a^2 + k_2.
$
Multiplying by $\vol$, we obtain the ODE
\begin{equation}
\frac{\mbox{d}\overline{A}}{\dt}
=
- 2 k_1/\vol \,\overline{A}^{\,2} + k_2 \vol
\label{ODEdimer2}
\end{equation}
where 
$
\overline{A}(t) = a(t) \vol 
$
is the deterministic approximation of the average number of 
molecules in the volume $\vol$ with concentration $a(t)$. 
Notice that equation (\ref{ODEdimer2}) does not give 
us the time evolution of the stochastic mean. To see that,
let us consider the stationary value
of $\overline{A}(t)$. It is given as the solution
of the stationary equation corresponding
to (\ref{ODEdimer2}), namely
$
0
=
- 2 k_1/\vol \overline{A}_s^{\,2} + k_2 \vol.
$
Hence, $\overline{A}_s = \vol \sqrt{k_2/2 k_1}$
which is not equal to $M_s$ given by formula (\ref{formulaMsGs}).
See also the discussion at the end of Section \ref{sechomoreactionexample}.

\section{Reaction-diffusion master equation}

\label{secRDMEmodsys}

Let $\en = \{0, 1, 2, 3, \dots\}$ be the set of non-negative
integers. Let $\boldn \in \en^{I_{all}}$ and $\boldm \in \en^{I_{all}}$.
We denote their coordinates by three indices, namely
\begin{equation}
\boldn= \big\{ n_{ijk} \, | \, (i,j,k) \in I_{all} \big\}
\qquad \mbox{and} \qquad
\boldm= \big\{ m_{ijk} \, | \, (i,j,k) \in I_{all} \big\}.
\label{tensorcoord}
\end{equation}
Let $p(\boldn,\boldm,t)$ be the joint probability that 
$A_{ijk}(t)=n_{ijk}$ and $B_{ijk}(t)=m_{ijk}$ for
all $(i,j,k) \in I_{all}$. The reaction-diffusion master 
equation describes the time evolution of $p(\boldn,\boldm,t)$.
To formulate it, we define the operators 
$
J_{ijk}^\bolde \, : \, \en^{I_{all}} \to \en^{I_{all}}
$
for
$
(i,j,k) \in I_{all}
$
and
$
\bolde \in \boldE_{ijk}
$
by
\begin{equation*}
J_{ijk}^\bolde (\boldn)
=
\big\{ q_{uvw} \, | \, (u,v,w) \in I_{all} \big\}
\end{equation*}
where
\begin{equation*}
q_{uvw}
=
\left\{
\begin{array}{ll}
n_{uvw}+1, \quad & \mbox{for} \; (u,v,w)=(i,j,k); \\
n_{uvw}-1, \quad & \mbox{for} \; (u,v,w)=(i,j,k) + \bolde; \\
n_{uvw}, \quad & \mbox{otherwise}.
\end{array}
\right.
\end{equation*}
We also define
$$
\bolddelta_{ijk}
=
\big\{ \delta_{uvw} \, | \, (u,v,w) \in I_{all} \big\}
\quad\;
\mbox{where}
\;\;\;
\delta_{uvw}
=
\left\{
\begin{array}{ll}
1, \;\; & \mbox{for} \; (u,v,w)=(i,j,k); \\
0, \;\; & \mbox{otherwise}.
\end{array}
\right.
$$
Then the reaction-diffusion master equation, i.e. the
chemical master equation
which corresponds to the system of ``chemical reactions" 
(\ref{abexamplecompartments})--(\ref{diffusionjumpsB}), 
can be written as follows \cite{Gillespie:1992:MPI,Gillespie:2000:CLE}
\begin{eqnarray}
\fl \qquad \quad \;
\frac{\partial p(\boldn,\boldm)}{\partial t} 
& = & 
\frac{k_1}{h^3}
\sum_{(i,j,k) \in I_{all}}
\Big\{
(n_{ijk}+1) m_{ijk}
 \, p (\boldn + \bolddelta_{ijk},\boldm)
- 
n_{ijk} m_{ijk} \, p (\boldn,\boldm)
\Big\}
\nonumber
\\
& + &
k_2 h^3
\sum_{(i,j,k) \in I_{all}}
\Big\{
p (\boldn - \bolddelta_{ijk},\boldm)
- 
p (\boldn,\boldm)
\Big\}
\label{cmeRD}
\\
& + & 
\frac{D_A}{h^2} \;
\sum_{(i,j,k) \in I_{all}}
\sum_{\bolde \in \boldE_{ijk}}
\Big\{
(n_{ijk}+1)
 \, p ( J_{ijk}^\bolde(\boldn),\boldm )
- 
n_{ijk} \, p (\boldn,\boldm)
\Big\}
\nonumber
\\
& + &
\frac{D_B}{h^2} \;
\sum_{(i,j,k) \in I_{all}}
\sum_{\bolde \in \boldE_{ijk}}
\Big\{
(m_{ijk}+1)
 \, p ( \boldn,J_{ijk}^\bolde(\boldm) )
- 
m_{ijk} \, p (\boldn,\boldm)
\Big\}.
\nonumber
\end{eqnarray}%
The first two terms on the right hand side correspond
to chemical reactions (\ref{abexamplecompartments}),
the third term to diffusion jumps (\ref{diffusionjumpsA})
and the last term to diffusion jumps (\ref{diffusionjumpsB}). 
 
\section{Derivation of formulae (\ref{propfunc2}) and (\ref{betaformula})}

\label{appenC}

We will first study the case $D_B=0$ and $B_0 = 1$. This means that
there is only one molecule of $B$ in the system and it does not
diffuse. In particular, reactions (\ref{diffusionjumpsB})
are not included in the model. Let the molecule of $B$ be
in the compartment $\cn = (b_1,b_2,b_3) \in I_{all}$.
Let $\boldn \in \en^{I_{all}}$ with the coordinates defined
by (\ref{tensorcoord}). Let $p(\boldn,t)$ be the joint 
probability that $A_{ijk}(t)=n_{ijk}$ for all
$(i,j,k) \in I_{all}$. Since the position of the molecule
of $B$ does not evolve, the reaction-diffusion master
equation (\ref{cmeRD}) simplifies to the following
equation for $p(\boldn,t)$
\begin{equation*}
\fl \qquad
\frac{\partial p(\boldn)}{\partial t} 
=
\frac{k_1}{h^3}
\Big\{
(n_{\cn}+1)
 \, p (\boldn + \bolddelta_{\cn})
- 
n_{\cn} \, p (\boldn)
\Big\}
+
k_2 h^3 \!\!
\sum_{(i,j,k) \in I_{all}}
\Big\{
p (\boldn - \bolddelta_{ijk})
- 
p (\boldn)
\Big\} 
\end{equation*} 
\begin{equation}
+ \;
\frac{D_A}{h^2} \;
\sum_{(i,j,k) \in I_{all}}
\sum_{\bolde \in \boldE_{ijk}}
\Big\{
(n_{ijk}+1)
 \, p ( J_{ijk}^\bolde(\boldn))
- 
n_{ijk} \, p (\boldn)
\Big\}.\qquad
\label{RDMEdbiszero}
\end{equation}
We want to change the reaction rate $k_1/h^3$ (of the bimolecular 
reaction per one compartment) to a reaction rate $\lambda$ in order 
to decrease the error between stationary distributions $\phi_K$ 
and $\phi_1$. The stationary version of (\ref{RDMEdbiszero}) 
with $k_1/h^3$ replaced by $\lambda$ is
\begin{eqnarray}
&&
\lambda
\Big\{
(n_{\cn}+1)
 \, p_s (\boldn + \bolddelta_{\cn})
- 
n_{\cn} \, p_s (\boldn)
\Big\}
+
k_2 h^3 \!\!
\sum_{(i,j,k) \in I_{all}}
\Big\{
p_s (\boldn - \bolddelta_{ijk})
- 
p_s (\boldn)
\Big\} 
\nonumber
\\
&& + \;
\frac{D_A}{h^2} \;
\sum_{(i,j,k) \in I_{all}}
\sum_{\bolde \in \boldE_{ijk}}
\Big\{
(n_{ijk}+1)
 \, p_s ( J_{ijk}^\bolde(\boldn))
- 
n_{ijk} \, p_s (\boldn)
\Big\} \; = \; 0\qquad
\label{RDMEdbiszerostat}
\end{eqnarray}
where
\begin{equation}
p_s (\boldn) = \lim_{t \to \infty} p(\boldn,t).
\label{defpsnapp}
\end{equation}
Let us denote the average number of molecules at the lattice site
$(i,j,k)$ as 
$$
M_{ijk}(t) 
= \sum_{\boldn} n_{ijk} \, p_s (\boldn)
\equiv
\sum_{n_{000}=0}^{\infty} \; \sum_{n_{001}=0}^{\infty} \cdots
\sum_{n_{KKK}=0}^{\infty} n_{ijk} \, p_s (\boldn).
$$
Multiplying (\ref{RDMEdbiszerostat}) by $n_{ijk}$ and summing 
over $\boldn$, we obtain
\begin{eqnarray}
k_2 h^3
+
\frac{D_A}{h^2} \;
\sum_{\bolde \in {\boldE}_{ijk}}
(M_{ijk+\bolde}
-
M_{ijk})
& = & 
0,
\qquad
\mbox{for} \; (i,j,k) \ne \cn, 
\label{disMboun1}
\\
k_2 h^3
+
\frac{D_A}{h^2} \;
\sum_{\bolde \in {\boldE}_{\cn}}
(M_{\cn + \bolde}
-
M_{\cn})
& = & 
\lambda
M_{\cn}.
\label{disMboun2}
\end{eqnarray}
Let us define tensors 
$\psi^{i^\prime j^\prime k^\prime} \in \er^{K \times K \times K}$,
for $i^\prime = 0, 1, 2, \dots, K-1$; $j^\prime = 0, 1, 2, \dots, K-1$
and $k^\prime = 0, 1, 2, \dots, K-1$, by
\begin{equation*}
\fl
\psi^{i^\prime j^\prime k^\prime}_{ijk}
=
\frac{1}{K^{3/2}}
\,
\cos \left( \frac{i^\prime (i-1/2) \pi}{K} \right)
\times 
\cos \left( \frac{j^\prime (j-1/2) \pi}{K} \right)
\times 
\cos \left( \frac{k^\prime (k-1/2) \pi}{K} \right)
\end{equation*}
\begin{equation}
\times \,
\left\{
\begin{array}{ll}
\sqrt{8}, & \quad \mbox{if $i^\prime$, $j^\prime$, $k^\prime$ 
are nonzero;} \; \\
\sqrt{4}, & \quad \mbox{if exactly one of 
$i^\prime$, $j^\prime$, $k^\prime$ is zero;} \; \\
\sqrt{2}, & \quad \mbox{if exactly two of 
$i^\prime$, $j^\prime$, $k^\prime$ are zero;} \; \\ 
1, & \quad \mbox{for $i^\prime = j^\prime = k^\prime = 0$.} \; 
\end{array}
\right. 
\label{defpsiFour}
\end{equation}
We have
\begin{equation*}
\sum_{i=1}^{K}
\cos \left( \frac{i^\prime (i-1/2) \pi}{K} \right)
\cos \left( \frac{i^{\prime\prime} (i-1/2) \pi}{K} \right)
= 
\delta_{i^\prime i^{\prime\prime}}
\frac{K}{2}, \qquad \mbox{for} \; i^\prime>0,
\end{equation*}
where $\delta_{i^\prime i^{\prime\prime}}$ is the Kronecker delta. 
Consequently, $\psi^{i^\prime j^\prime k^\prime}$, for 
$i^\prime, j^\prime, k^\prime = 0, 1, \dots, K-1$, 
satisfy the orthonormality condition:
\begin{equation}
\sum_{i,j,k=1}^{K}
\psi^{i^\prime j^\prime k^\prime}_{ijk} 
\psi^{i^{\prime\prime} j^{\prime\prime} k^{\prime\prime}}_{ijk}
=
\delta_{i^\prime i^{\prime\prime}}
\delta_{j^\prime j^{\prime\prime}}
\delta_{k^\prime k^{\prime\prime}}.
\label{orthornompsi}
\end{equation}
Let us express $M_{ijk}$ in the basis $\psi^{i^\prime j^\prime k^\prime}$:
\begin{equation}
M_{ijk}
=
\sum_{i^\prime, j^\prime, k^\prime=0}^{K-1} 
\tildeM_{i^\prime j^\prime k^\prime} 
\, 
\psi^{i^\prime j^\prime k^\prime}_{ijk}.
\label{disfourtra}
\end{equation}
Then (\ref{disMboun1})--(\ref{disMboun2}) read as follows
\begin{eqnarray*}
k_2 h^3
+
\frac{D_A}{h^2}
\sum_{i^\prime, j^\prime, k^\prime=0}^{K-1} 
\tildeM_{i^\prime j^\prime k^\prime} 
\sum_{\bolde \in {\boldE}_{ijk}}
\left(
\psi^{i^\prime j^\prime k^\prime}_{ijk+\bolde}
-
\psi^{i^\prime j^\prime k^\prime}_{ijk}
\right)
& = & 
0,
\quad
\mbox{for} \; (i,j,k) \ne \cn, 
\\
k_2 h^3
+
\frac{D_A}{h^2} 
\sum_{i^\prime, j^\prime, k^\prime=0}^{K-1} 
\tildeM_{i^\prime j^\prime k^\prime} 
\sum_{\bolde \in {\boldE}_{\cn}}
\left(
\psi^{i^\prime j^\prime k^\prime}_{\cn+\bolde}
-
\psi^{i^\prime j^\prime k^\prime}_{\cn}
\right)
& = & 
\lambda
M_{\cn}.
\end{eqnarray*}
Using (\ref{defpsiFour}), we get
\begin{eqnarray}
k_2 h^3
+
\frac{D_A}{h^2}
\sum_{i^\prime, j^\prime, k^\prime=0}^{K-1} 
\tildeM_{i^\prime j^\prime k^\prime} \,
c^{i^\prime j^\prime k^\prime} \, 
\psi^{i^\prime j^\prime k^\prime}_{ijk}
& = & 
0,
\quad
\mbox{for} \; (i,j,k) \ne \cn, 
\label{pomdiscfourboun1}
\\
k_2 h^3
+
\frac{D_A}{h^2} 
\sum_{i^\prime, j^\prime, k^\prime=0}^{K-1} 
\tildeM_{i^\prime j^\prime k^\prime} \,
c^{i^\prime j^\prime k^\prime} \,
\psi^{i^\prime j^\prime k^\prime}_{\cn}
& = & 
\lambda
M_{\cn},
\label{pomdiscfourboun2}
\end{eqnarray}
where
\begin{equation}
c^{ijk}
=
2
\left(
\cos \left( \frac{i \pi}{K} \right)
+
\cos \left( \frac{j \pi}{K} \right)
+
\cos \left( \frac{k \pi}{K} \right)
- 3 
\right). 
\label{defbIJL}
\end{equation}
Multiplying (\ref{pomdiscfourboun1})--(\ref{pomdiscfourboun2}) by
$\psi^{i^{\prime\prime} j^{\prime\prime} k^{\prime\prime}}_{ijk}$, 
summing the resulting equations and using the orthonormality 
condition (\ref{orthornompsi}), we obtain
\begin{eqnarray}
\qquad\qquad\quad\;\;\;
k_2 \vol 
&=& 
\lambda
M_{\cn},
\label{pom1derbeta}
\\
\frac{D_A}{h^2}
\, c^{i^{\prime\prime}j^{\prime\prime}k^{\prime\prime}}
\tildeM_{i^{\prime\prime}j^{\prime\prime}k^{\prime\prime}}
&=& 
\lambda
M_{\cn}
\,
\psi^{i^{\prime\prime}j^{\prime\prime}k^{\prime\prime}}_{\cn},
\qquad
\mbox{for} \; (i^{\prime\prime},j^{\prime\prime},k^{\prime\prime}) \ne (0,0,0),
\nonumber
\end{eqnarray}
where $\vol = L^3 = h^3 K^3$ is the volume of the reactor.
We drop the double primes on indices $i$, $j$ and $k$ 
to simplify the notation and obtain
\begin{equation*}
\tildeM_{ijk}  
= 
\frac{\lambda \, h^2 M_{\cn} 
\, \psi^{ijk}_{\cn}}{D_A \, c^{ijk}},
\qquad
\mbox{for} \; (i, j, k) \ne (0,0,0).
\end{equation*}
Using (\ref{pom1derbeta}), we get
\begin{equation}
\tildeM_{ijk}  
= 
\frac{k_2 \vol \, h^2 \, \psi^{ijk}_{\cn}}{D_A \, c^{ijk}},
\qquad
\mbox{for} \; (i, j, k) \ne (0,0,0).
\label{pom2derbeta}
\end{equation}
Using (\ref{disfourtra}) and (\ref{pom2derbeta}), we have
\begin{equation}
M_{\cn}
=
\sum_{i,j,k=0}^{K-1} 
\tildeM_{ijk} \, \psi^{ijk}_{\cn}
=
K^{-3/2} \, \tildeM_{000}
-  
\frac{k_2 \vol \, h^2}{D_A}
\beta_{\cn}
\label{pom3derbeta}
\end{equation}
where
\begin{equation}
\beta_{\cn}
=
- \!\!\!
\sum_{\hbox{\hsize=0.145\hsize\vbox{
      \noindent
      \scriptsize
      \;\;\;\;\;$i,j,k=0$ \\
      $(i,j,k) \ne (0,0,0)$}}}^{K-1} 
\frac{(\psi^{ijk}_{\cn})^2}{c^{ijk}}.
\label{betabdef}
\end{equation}
Substituting (\ref{pom1derbeta}) into (\ref{pom3derbeta}), we have
\begin{equation}
\frac{k_2 \vol}{\lambda}
=
K^{-3/2} \, \tildeM_{000}
-  
\beta_{\cn}
\frac{k_2 \vol \, h^2}{D_A}.
\label{fourcoefboun}
\end{equation}
The average number of molecules of $A$ in the reactor is given by
\begin{equation*}
A_s
\equiv
\sum_{i,j,k=1}^{K}
M_{ijk}.
\end{equation*}
Using (\ref{disfourtra}), we get
\begin{equation}
\fl
A_s
=
\sum_{i,j,k=1}^{K}
\sum_{i^\prime, j^\prime, k^\prime=0}^{K-1} 
\tildeM_{i^\prime j^\prime k^\prime} 
\, 
\psi^{i^\prime j^\prime k^\prime}_{ijk}
=
\sum_{i^\prime, j^\prime, k^\prime=0}^{K-1} 
\tildeM_{i^\prime j^\prime k^\prime} 
\!\!
\sum_{i,j,k=1}^{K}
\psi^{i^\prime j^\prime k^\prime}_{ijk}
=
\tildeM_{000} \, K^{3/2}.
\label{pomaverA}
\end{equation}
We would like to choose $\lambda$ so that $A_s = M_s$ where
$M_s$ is given by (\ref{meanMsabexample}), i.e.
\begin{equation*}
\frac{k_2 \vol^2}{k_1}
=
\tildeM_{000} \, K^{3/2}.
\end{equation*}
Substituting for $\tildeM_{000}$ into (\ref{fourcoefboun}) and using
$\vol = h^3 K^3$, we get
\begin{equation*}
\frac{k_2 \vol}{\lambda}
=
h^3 \, \frac{k_2 \vol}{k_1}
-  
\beta_{\cn} \,
\frac{k_2 \vol\, h^2}{D_A}
\end{equation*}
which implies
\begin{equation*}
\lambda
=
\frac{D_A k_1}
{D_A h^3 - \beta_{\cn} \, k_1 h^2}.
\end{equation*}
This choice of $\lambda$ gives the average number of molecules of
$A$ equal to $M_s$, provided that $D_B = 0$, $B_0 = 1$ and the 
molecule of $B$ is in the compartment $\cn = (b_1,b_2,b_3) \in I_{all}$.

Now let $D_B \ne 0$ and $B_0 = 1$. If we want to model the bimolecular reaction 
(\ref{abexample}), it is important to know the distances of molecules of 
$A$ from the molecule of $B$. The distances diffuse with the diffusion 
constant $D_A+D_B$. Thus we can equivalently model their time evolution 
by considering that the molecule of $B$ does not diffuse and molecules of
$A$ diffuse with the diffusion constant $D_A + D_B$. Then the
previous calculation (equation (\ref{fourcoefboun})) implies that
\begin{equation}
K^{-3/2} \, \tildeM_{000}^{\cn}
=  
\frac{k_2 \vol}{\lambda}
+
\beta_{\cn} \,
\frac{k_2 \vol \, h^2}{D_A + D_B}
\label{fourcoefboun2}
\end{equation}
where notation $\tildeM_{000}^{\cn} \equiv \tildeM_{000}$ highlights 
the fact that $\tildeM_{000}$ depends on the position $\cn$ of the molecule
of $B$. Let $p_{\cn}$ be the probability that the
molecule of $B$ is in the compartment $\cn = (b_1,b_2,b_3) \in I_{all}$.
We have $p_{\cn} = K^{-3}$. The average number of molecules of $A$ in 
the reactor is given by (compare with (\ref{pomaverA}))
\begin{equation*}
A_s
\equiv
\sum_{\cn}
p_{\cn} \,
\tildeM_{000}^{\cn} \, K^{3/2}
=
K^{-3/2}
\sum_{\cn}
\tildeM_{000}^{\cn}.
\end{equation*}
Multyplying (\ref{fourcoefboun2}) by $p_{\cn} = K^{-3}$ and
summing over $\cn$, we obtain
\begin{equation}
\frac{A_s}{K^3}
=
\frac{k_2 \vol}{\lambda}
+
\beta \,
\frac{k_2 \vol \, h^2}{D_A + D_B}
\label{betapom1} 
\end{equation}
where 
\begin{equation}
\beta
=
\sum_{\cn}
p_{\beta} \, \beta_{\cn} 
=
\frac{1}{K^3}
\sum_{\cn}
\beta_{\cn}.
\label{betapom2} 
\end{equation}
We would like to choose $\lambda$ so that $A_s = M_s$ where
$M_s$ is given by (\ref{meanMsabexample}). Substituting
(\ref{meanMsabexample}) into (\ref{betapom1}), we get 
\begin{equation*}
\frac{k_2 \vol^2}{k_1 K^3}
=
\frac{k_2 \vol}{\lambda}
+
\beta \,
\frac{k_2 \vol \, h^2}{D_A + D_B}.
\end{equation*}
Using $\vol = h^3 K^3$, we obtain
\begin{equation*}
\lambda
=
\frac{(D_A + D_B) k_1}{(D_A + D_B) h^3 - \beta k_1 h^2}.
\end{equation*}
Thus we have derived (\ref{propfunc2}). Using (\ref{betapom2}),
(\ref{betabdef}) and 
orthonormality condition (\ref{orthornompsi}), we obtain
$$
\beta
=
\frac{1}{K^3}
\sum_{\cn}
\beta_{\cn}
=
-
\frac{1}{K^3} \!\!\!
\sum_{\hbox{\hsize=0.145\hsize\vbox{
      \noindent
      \scriptsize
      \;\;\;\;\;$i,j,k=0$ \\
      $(i,j,k) \ne (0,0,0)$}}}^{K-1} 
\sum_{\cn}
\frac{(\psi^{ijk}_{\cn})^2}{c^{ijk}}
=
-
\frac{1}{K^3} \!\!\!
\sum_{\hbox{\hsize=0.145\hsize\vbox{
      \noindent
      \scriptsize
      \;\;\;\;\;$i,j,k=0$ \\
      $(i,j,k) \ne (0,0,0)$}}}^{K-1} 
\frac{1}{c^{ijk}}.
$$
Substituting (\ref{defbIJL}) for $c^{ijk}$, we obtain
(\ref{betaformula}).

\section{Derivation of formula (\ref{betainfinityformula})}

\label{appdercomp}

Formula (\ref{betaformula}) is the Riemann sum of the definite integral
\begin{equation}
\beta_\infty
=
\frac{1}{2(\pi)^3}
\int_0^{\pi}
\int_0^{\pi}
\int_0^{\pi}
\frac{1}{3 - \cos x - \cos y - \cos z} \,
\mbox{d}x \,
\mbox{d}y \,
\mbox{d}z,
\label{betainfinitytripleintegral}
\end{equation}
i.e., passing $K \to \infty$ in (\ref{betaformula}), we obtain 
(\ref{betainfinitytripleintegral}). Integrating over $z$, we get
(\ref{betainfinityformula}). 

Formula (\ref{betainfinityformula}) can be also derived directly
without the help of (\ref{betaformula}). Such a derivation uses
a similar reasoning as the derivation of (\ref{formk1lambda})
in \ref{appderk1lambda}, i.e. it establishes a link between 
molecular-based models and the compartment-based modelling.
We consider the infinite three-dimensional lattice
\begin{equation}
(i,j,k) h
\qquad
\mbox{for}
\;
i \in \zet,
\;
j \in \zet,
\;
k \in \zet
\label{infinitelattice}
\end{equation}
where $h \in \er$. To model bimolecular reactions by the 
compartment-based approach, we need to know whether the molecules 
are in the same compartment or not. In particular, it is sufficient 
to track the relative distance of molecules rather than their absolute 
positions. Postulating that the molecule of $B$ is always at the 
origin (compartment $(0,0,0)$) and letting the molecule of $A$ 
diffuse with the diffusion constant $(D_A+D_B)$, we obtain
the stochastic model which gives the same distribution
of relative distances of molecules as the original stochastic
model. Thus we will study the following auxiliary stochastic
process. We consider that the particles jump 
to neighbouring lattice sites with the rate $(D_A+D_B)/h^2$ and
are removed at the origin with the rate $\lambda$. 
The reaction-diffusion master equation can be written for
this model as follows (using the same notation as in (\ref{cmeRD}))
\begin{eqnarray}
\frac{\partial p(\boldn)}{\partial t} 
& = & 
\frac{D_A+D_B}{h^2} \;
\sum_{(i,j,k) \in \zet^3}
\sum_{\bolde \in \boldE}
\Big\{
(n_{ijk}+1)
 \, p ( J_{ijk}^\bolde(\boldn) )
- 
n_{ijk} \, p (\boldn)
\Big\}
\nonumber
\\
& + &
\lambda
\Big\{
(n_{000}+1)
 \, p (\boldn + \bolddelta_{000})
- 
n_{000} \, p (\boldn)
\Big\}.
\label{cmeRDlambdah}
\end{eqnarray}%
We are interested in the stationary behaviour of a system of 
(infinitely) many molecules of $A$, subject to the condition that 
the average number of molecules per compartment is kept constant 
(equal to $M_\infty$) far from the origin, i.e. in the limit 
$\sqrt{i^2+j^2+k^2} \to \infty.$ The stationary version of 
(\ref{cmeRDlambdah}) reads as follows
\begin{equation*}
\frac{D_A+D_B}{h^2} \;
\sum_{(i,j,k) \in \zet^3}
\sum_{\bolde \in \boldE}
\Big\{
(n_{ijk}+1)
 \, p_s ( J_{ijk}^\bolde(\boldn) )
- 
n_{ijk} \, p_s(\boldn)
\Big\} 
\end{equation*}
\begin{equation}
=
- 
\lambda
\Big\{
(n_{000}+1)
 \, p_s (\boldn + \bolddelta_{000})
- 
n_{000} \, p_s (\boldn)
\Big\}
\label{stationaryinfinRDME}
\end{equation}
where $p_s(\boldn)$ is defined as in (\ref{defpsnapp}).
Let us denote the average number of molecules at the lattice site
$(i,j,k)$ as 
$$
M_{ijk}(t) = \sum_{\boldn} n_{ijk} p_s (\boldn).
$$
Multiplying (\ref{stationaryinfinRDME}) by $n_{ijk}$ and summing 
over $\boldn$, we obtain
\begin{eqnarray*}
\frac{D_A+D_B}{h^2} \;
\sum_{\bolde \in \boldE}
(M_{ijk+\bolde}
-
M_{ijk})
& = & 
0
\qquad
\mbox{for} \; (i,j,k) \ne (0,0,0), \\
\frac{D_A+D_B}{h^2} \;
\sum_{\bolde \in \boldE}
(M_{\bolde}
-
M_{000})
& = & 
\lambda
M_{000}.
\end{eqnarray*}
Let us define $\mu_{ijk} = M_{ijk} - M_\infty$. Then we have
\begin{eqnarray*}
\sum_{\bolde \in \boldE}
\mu_{ijk+\bolde}
& = & 
6 \mu_{ijk},
\qquad
\mbox{for} \; (i,j,k) \ne (0,0,0), 
\\
\sum_{\bolde \in \boldE}
\mu_{\bolde}
& = & 
6 \mu_{000}
+ 
\frac{\lambda \, h^2}{D_A+D_B} \left( \mu_{000} + M_\infty \right).
\end{eqnarray*}
Multiplying by 
$\exp^{\imathmys \, x i} \exp^{\imathmys \, y j} \exp^{\imathmys \, z k}$,
where $\imathmy = \sqrt{-1}$, 
$x \in \er$, $y \in \er$, $z \in \er$, 
and summing over $i,$ $j$ and $k$, we obtain
\begin{eqnarray}
6 \, \widehat{\mu}_{xyz}
& = &
\widehat{\mu}_{xyz} 
\left( \exp^{\imathmys \, x} + \exp^{- \imathmys \, x}
+ \exp^{\imathmys \, y} + \exp^{- \imathmys \, y}
+ \exp^{\imathmys \, z} + \exp^{- \imathmys \, z} \right)
\nonumber
\\
&-&
\lambda \, h^2 \left( \mu_{000} + M_\infty \right)/(D_A+D_B)
\label{pomfourinf}
\end{eqnarray}
where $\widehat{\mu}_{xyz}$ is the 
Fourier transform
\begin{equation*}
\widehat{\mu}_{xyz}
= 
\sum_{i=-\infty}^\infty 
\,
\sum_{j=-\infty}^\infty 
\,
\sum_{k=-\infty}^\infty 
\exp^{\imathmys \, x i} \exp^{\imathmys \, y j} \exp^{\imathmys \, z k} 
\mu_{ijk}.
\end{equation*}
Simplifying (\ref{pomfourinf}), we obtain 
\begin{equation*}
\widehat{\mu}_{xyz}
=
\frac{\lambda \, h^2 \left( \mu_{000} + M_\infty \right)}{2 (D_A+D_B)} 
\frac{1}{\cos x + \cos y + \cos z - 3}.
\end{equation*}
Thus
\begin{equation}
\mu_{000}
=
- \beta_\infty
 \frac{\lambda \, h^2 \left( \mu_{000} + M_\infty \right)}{D_A+D_B}
\label{eqmuzero}
\end{equation}
where $\beta_\infty$ is the constant given by
\begin{equation}
\beta_\infty 
=
\frac{1}{2 (2\pi)^3}
\int_0^{2 \pi}
\int_0^{2 \pi}
\int_0^{2 \pi}
\frac{1}{3 - \cos x - \cos y - \cos z} \;
\mbox{d}x \,
\mbox{d}y \,
\mbox{d}z.
\label{betainfty2pi}
\end{equation}
Using $\mu_{000} = M_{000} - M_\infty$, equation (\ref{eqmuzero})
can be rewritten as
\begin{equation}
\lambda M_{000}
=
\frac{\lambda \, (D_A+D_B)}{(D_A+D_B) + \lambda \, \beta_\infty h^2} \,
M_\infty.
\label{rateofremovalofA}
\end{equation}
The term $\lambda M_{000}$ gives the rate of removal of molecules
of $A$ at the origin. The rate of change of the concentration
$a$ of molecules of $A$, which is subject to heteroreaction 
(\ref{abexample}), can be also described by the deterministic ODE
$da/dt = - k_1 a b$ where $b$ is the concentration of
molecules of $B$. This ODE can be equivalently rewritten in terms
of the average numbers of molecules of $A$ and $B$ per lattice
site, i.e. in terms of $A=a h^3$ and $B = b h^3$, as
$dA/dt = - k_1/h^3 A B$. Using $B=1$ and $A=M_\infty$, the
rate of removal of molecules of $A$ is given by
$k_1/h^3 M_\infty$. Comparing with (\ref{rateofremovalofA}),
we obtain
\begin{equation*}
\frac{k_1}{h^3} \, M_\infty 
= 
\frac{\lambda \, (D_A+D_B)}{(D_A+D_B) + \lambda \, \beta_\infty h^2} \,
M_\infty.
\end{equation*}
Solving for $\lambda$, we obtain
\begin{equation*}
\lambda
=
\frac{(D_A+D_B) k_1}{(D_A+D_B) h^3 - \beta_\infty k_1 h^2}.
\end{equation*}
Thus we have derived (\ref{propfunc2}) with $\beta = \beta_\infty$.
The constant $\beta_\infty$ is given by (\ref{betainfty2pi}).
Using periodicity of the cosine function, we obtain
(\ref{betainfinitytripleintegral}). Integrating over $z$, we 
derive (\ref{betainfinityformula}). 

\section{Derivation of formula (\ref{formk1lambda})}

\label{appderk1lambda}

In order to derive (\ref{formk1lambda}), we consider
the diffusion to the ball of radius $\newrho$ which removes 
molecules of $A$ with the rate $\lambda$. Let the centre 
of the ball be at the origin. Let $c(r)$ be the equilibrium
concentration of molecules of $A$ at distance $r$ from the 
origin, which is a continuous function with continuous derivative 
satisfying the equations
\begin{eqnarray*}
\frac{\mbox{d}^2 c}{\mbox{d} r^2}
+
\frac{2}{r}
\frac{\mbox{d} c}{\mbox{d} r} 
& = & 0, \qquad \mbox{for} \; r \ge \newrho,
\\
\frac{\mbox{d}^2 c}{\mbox{d} r^2}
+
\frac{2}{r}
\frac{\mbox{d} c}{\mbox{d} r} 
- \frac{\lambda \, c}{D_A+D_B} 
& = & 0, \qquad \mbox{for} \; r \le \newrho.
\end{eqnarray*}
The general solution of these second-order 
ODEs can be written
in the following form
\begin{eqnarray*}
\fl \qquad c(r) 
& = & 
a_1 + \frac{a_2}{r}, \qquad
\qquad 
\qquad
\qquad
\qquad  
\qquad
\qquad
\qquad
\qquad
\qquad 
\;
\mbox{for} \; r \ge \newrho,
\\
\fl \qquad c(r) 
& = & 
\frac{a_3}{r} \, \exp\left[ r \, \sqrt{\frac{\lambda}{D_A+D_B}} \right]
+
\frac{a_4}{r} \, \exp\left[ - r \, \sqrt{\frac{\lambda}{D_A+D_B}} \right], 
\qquad \mbox{for} \; r \le \newrho,
\end{eqnarray*}
where $a_1$, $a_2$, $a_3$ and $a_4$ are real constants.
We impose the boundary condition at infinity
\begin{equation*}
\lim_{r \to \infty}
c(r)
=
c_{\infty}.
\end{equation*}
This implies $a_1 = c_\infty$. Since $c$ is continuous at the origin,
we deduce $a_4 = - a_3$. Thus we have
\begin{eqnarray*}
c(r) 
& 
=
&  
c_{\infty} + \frac{a_2}{r}, 
\qquad 
\qquad
\qquad
\qquad 
\qquad
\qquad
\qquad
\qquad
\quad
\mbox{for} \; r \ge \newrho,
\\
c(r) 
&
=
& 
\frac{2 a_3}{r} \, 
\sinh \left( r \, \sqrt{\frac{\lambda}{D_A+D_B}} \right), 
\qquad
\qquad
\qquad
\qquad
\quad
\,
\mbox{for} \; r \le \newrho.
\end{eqnarray*}
To determine the constants $a_2$ and $a_3$, we use the continuity
of $c$ and its derivative at $r = \newrho$. We obtain
\begin{eqnarray*}
a_2
& = &
c_\infty
\left\{
\sqrt{(D_A+D_B)/\lambda} \, 
\tanh \left( \newrho \, \sqrt{\lambda/(D_A+D_B)} \right)
- 
\newrho
\right\},
\\
a_3 
& = &
c_\infty \, \sqrt{(D_A+D_B)/\lambda} \, 
\left( 2 \, \cosh \left( \newrho \, \sqrt{\lambda/(D_A+D_B)} \right)
\right)^{-1}.
\end{eqnarray*}
The flux through the unit area of the boundary can be computed as
\begin{equation*}
\left.
(D_A+D_B) \frac{\partial c}{\partial r} 
\right|_{r=\newrho}
=
- 
\frac{(D_A+D_B) a_2}{\newrho^2}.
\end{equation*}
The area of the sphere is $4 \pi \newrho^2$. Thus the total flux
through the sphere boundary is $- 4 \pi (D_A+D_B) a_2$. Substituting
for $a_2$, we get
\begin{equation*}
4 \pi (D_A+D_B)
\left(
\newrho
-
\sqrt{(D_A+D_B)/\lambda} \, 
\tanh \left( \newrho \, \sqrt{\lambda/(D_A+D_B)} \right)
\right)
c_\infty.
\end{equation*}
This quantity is equal to the rate constant of bimolecular reaction
$k_1$ multiplied by the concentration of the chemical far from the reacting
molecule $c_\infty$. Dividing by $c_\infty$, we derive (\ref{formk1lambda}).
Let us note that we used diffusion to the ball to derive
(\ref{formk1lambda}). This approximation can be justified using
the more general evolution equation for the many particle distribution 
function \cite{Doi:1976:STD}. 

\section{Derivation of (\ref{IntegrEquation2})--(\ref{imporequation})
and a numerical method for solving it}

\label{appendF}

Let $c_i(r)$ be the concentration of molecules of $A$ at distance
$r$ from the origin. Assuming that molecules of $A$ only diffuse, 
their concentration at point $r$ after the time interval $\Delta t$ 
is given as
\begin{equation}
\int_0^\infty 
K(r,r^\prime; \gamma) \,
c_i(r^\prime)
\, \dr^\prime
\label{cfdiffusion}
\end{equation}
where $K(r,r^\prime; \gamma)$ is given by (\ref{defKrr}). Let us
assume that the particles are removed, in the circle of radius 
$\newrho$ and centered at origin, with probability $P_\lambda$, 
and then diffuse for time $\Delta t$. 
Then (\ref{cfdiffusion}) is modified to
\begin{equation*}
c_{i+1}(r)
=
(1 - P_\lambda)
\int_0^1
K(r,r^\prime; \gamma) \,
c_i(r^\prime)
\, \dr^\prime
+
\int_1^\infty 
K(r,r^\prime; \gamma) \,
c_i(r^\prime)
\, \dr^\prime.
\end{equation*}
Equation (\ref{IntegrEquation2}) is an equation for the fixed
point of this iterative scheme. The function $g(r)$ is the generalization
of the radial distribution function (RDF) for bimolecular reaction at
steady state \cite{Andrews:2004:SSC} for arbitrary 
$P_\lambda \in [0,1]$. Note that the RDF in \cite{Andrews:2004:SSC}
was only computed for $P_\lambda = 1$. The rate of removal of particles
(at steady state) during one time step is given by the right hand side 
of (\ref{imporequation}). Comparing with $\kappa$, we obtain 
(\ref{imporequation}).

To solve (\ref{IntegrEquation2}), we will use the condition
$g(r) \to 1$ as $r \to \infty$. Choosing $S$ large, we can
approximate $g(r) = 1$ for $r \ge S$. Let $N_1$ and $N_2$
be positive integers. We consider the mesh 
$r_j = j/N_1,$ for $j= 1, 2, \dots, N_1$ and
$r_j = 1 + (S - 1)(j-N_1)/N_2$, for $j= N_1 + 1, \dots, N_1 + N_2$.
We discretize (\ref{IntegrEquation2}) as
$$
g(r_i) 
=
\frac{1 - P_\lambda}{N_1}
\sum_{j=1}^{N}
K(r_i,r_j; \gamma)
g(r_j)
+
\frac{S - 1}{N_2}
\sum_{j=N_1}^{N_1 + N_2}
K(r_i,r_j; \gamma)
g(r_j)
+
\int_S^\infty
K(r_i,r^\prime; \gamma)
\, \dr^\prime.
$$
This is a linear system for $g(r_i)$, $i = 1, 2, \dots, N_1 + N_2$,
which can be solved, for example, by Gaussian elimination.
Let us note that the right hand side of this system can be evaluated
using the error function $\erf$ as
$$
\int_S^\infty
K(r_i,r^\prime; \gamma)
\, \dr^\prime
=
- \frac{\gamma^2 \, K(r_i,S)}{S}
+
1
-
\frac{1}{2} \, \erf \left[ \frac{S-r_i}{\gamma \sqrt{2}} \right]
- 
\frac{1}{2} \, \erf \left[ \frac{S+r_i}{\gamma \sqrt{2}} \right]. 
$$
Substituting $g(r_i)$, $i = 1, 2, \dots, N_1 + N_2$, into
(\ref{imporequation}), we compute $\kappa$. Repeating this
computation for different values of $\gamma$ and $P_\lambda$,
we obtain the results presented in Figure \ref{figdependencekPs}.

\section*{References}
\bibliographystyle{amsplain}
\bibliography{bibrad}

\providecommand{\bysame}{\leavevmode\hbox to3em{\hrulefill}\thinspace}
\providecommand{\MR}{\relax\ifhmode\unskip\space\fi MR }
\providecommand{\MRhref}[2]{%
  \href{http://www.ams.org/mathscinet-getitem?mr=#1}{#2}
}
\providecommand{\href}[2]{#2}
\begin{thebibliography}{10}

\bibitem{Alberts:2002:MBC}
B.~Alberts, A.~Johnson, J.~Lewis, M.~Raff, K.~Roberts, and P.~Walter,
  \emph{{M}olecular {B}iology of the {C}ell}, Garland Science, New York, 2002.

\bibitem{Andrews:2004:SSC}
S.~Andrews and D.~Bray, \emph{Stochastic simulation of chemical reactions with
  spatial resolution and single molecule detail}, Physical Biology \textbf{1}
  (2004), 137--151.

\bibitem{Berg:1983:RWB}
H.~Berg, \emph{{R}andom {W}alks in {B}iology}, Princeton University Press,
  1983.

\bibitem{Berg:1977:PC}
H.~Berg and E.~Purcell, \emph{Physics of chemoreception}, Biophysical Journal
  \textbf{20} (1977), 193--219.

\bibitem{Chandrasekhar:1943:SPP}
S.~Chandrasekhar, \emph{Stochastic problems in physics and astronomy}, Reviews
  of Modern Physics \textbf{15} (1943), 2--89.

\bibitem{DeVille:2006:NDL}
R.~DeVille, C.~Muratov, and E.~Vanden-Eijnden, \emph{Non-meanfield
  deterministic limits in chemical reaction kinetics}, Journal of Chemical
  Physics \textbf{124} (2006), 231102.

\bibitem{Doi:1976:STD}
M.~Doi, \emph{Stochastic theory of diffusion-controlled reaction}, Journal of
  Physics A: Mathematical and General \textbf{9} (1976), no.~9, 1479--1495.

\bibitem{Einstein:1905:UMT}
A.~Einstein, \emph{\"{U}ber die von der molekularkinetischen {T}heorie der
  {W}\"arme geforderte {B}ewegung von in ruhenden {F}l\"ussigkeiten
  suspendierten {T}eilchen}, Annalen der Physik \textbf{17} (1905), 549--560.

\bibitem{Engblom:2006:CMH}
S.~Engblom, \emph{Computing the moments of high dimensional solutions of the
  master equation}, Applied Mathematics and Computation \textbf{180} (2006),
  no.~2, 498--515.

\bibitem{Engblom:2008:SSR}
S.~Engblom, L.~Ferm, A.~Hellander, and P.~L\"otstedt, \emph{Simulation of
  stochastic reaction-diffusion processes on unstructured meshes}, Technical
  Report 2008-012, Dept of Information Technology, Uppsala University, Uppsala,
  Sweden, 2008.

\bibitem{Erban:2007:RBC}
R.~Erban and S.~J. Chapman, \emph{Reactive boundary conditions for stochastic
  simulations of reaction-diffusion processes}, Physical Biology \textbf{4}
  (2007), no.~1, 16--28.

\bibitem{Erban:2008:ASC}
R.~Erban, S.~J. Chapman, I.~Kevrekidis, and T.~Vejchodsky, \emph{Analysis of a
  stochastic chemical system close to a sniper bifurcation of its mean-field
  model}, submitted to SIAM Journal on Applied Mathematics, available as
  http://arxiv.org/abs/0807.4498, 2008.

\bibitem{Erban:2007:PGS}
R.~Erban, S.~J. Chapman, and P.~Maini, \emph{A practical guide to stochastic
  simulations of reaction-diffusion processes}, 35 pages, available as
  http://arxiv.org/abs/0704.1908, 2007.

\bibitem{Erban:2004:ICB}
R.~Erban and H.~Othmer, \emph{From individual to collective behaviour in
  bacterial chemotaxis}, SIAM Journal on Applied Mathematics \textbf{65}
  (2004), no.~2, 361--391.

\bibitem{Erban:2005:STS}
\bysame, \emph{From signal transduction to spatial pattern formation in {{\em
  E. coli}}: A paradigm for multi-scale modeling in biology}, Multiscale
  Modeling and Simulation \textbf{3} (2005), no.~2, 362--394.

\bibitem{Erban:2007:TEA}
\bysame, \emph{Taxis equations for amoeboid cells}, Journal of Mathematical
  Biology \textbf{54} (2007), no.~6, 847--885.

\bibitem{Gillespie:1977:ESS}
D.~Gillespie, \emph{Exact stochastic simulation of coupled chemical reactions},
  Journal of Physical Chemistry \textbf{81} (1977), no.~25, 2340--2361.

\bibitem{Gillespie:1992:MPI}
\bysame, \emph{{M}arkov {P}rocesses, an introduction for physical scientists},
  Academic Press, Inc., Harcourt Brace Jovanowich, 1992.

\bibitem{Gillespie:2000:CLE}
\bysame, \emph{The chemical {L}angevin equation}, Journal of Chemical Physics
  \textbf{113} (2000), no.~1, 297--306.

\bibitem{Hattne:2005:SRD}
J.~Hattne, D.~Fange, and J.~Elf, \emph{Stochastic reaction-diffusion simulation
  with {M}eso{RD}}, Bioinfor\-matics \textbf{21} (2005), no.~12, 2923--2924.

\bibitem{Isaacson:2008:RME}
S.~Isaacson, \emph{The reaction-diffusion master equation as an asymptotic
  approximation of diffusion to a small target}, to appear in SIAM Journal on
  Applied Mathematics, 2009.

\bibitem{Isaacson:2006:IDC}
S.~Isaacson and C.~Peskin, \emph{Incorporating diffusion in complex geometries
  into stochastic chemical kinetics simulations}, SIAM Journal on Scientific
  Computing \textbf{28} (2006), no.~1, 47--74.

\bibitem{McKane:2004:SMP}
A.~McKane and T.~Newman, \emph{Stochastic models in population biology and
  their deterministic analogs}, Physical Review E \textbf{70} (2004), 041902.

\bibitem{Murray:2002:MB}
J.~Murray, \emph{{M}athematical {B}iology}, Springer Verlag, 2002.

\bibitem{Paulsson:2000:SFF}
J.~Paulsson, O.~Berg, and M.~Ehrenberg, \emph{Stochastic focusing:
  {F}luctuation-enhanced sensitivity of intracellular regulation}, Proceedings
  of the National Academy of Sciences USA \textbf{97} (2000), no.~13,
  7148--7153.

\bibitem{Singer:2008:PRD}
A.~Singer, Z.~Schuss, A.~Osipov, and D.~Holcman, \emph{Partially reflected
  diffusion}, SIAM Journal on Applied Mathematics \textbf{68} (2008), no.~3,
  844--868.

\bibitem{Smoluchowski:1917:VMT}
M.~Smoluchowski, \emph{Versuch einer mathematischen {T}heorie der
  {K}oagulationskinetik kolloider {L}\"osungen}, Zeitschrift f\"ur
  physikalische Chemie \textbf{92} (1917), 129--168.

\bibitem{Stiles:2001:MCM}
J.~Stiles and T.~Bartol, \emph{{M}onte {C}arlo methods for simulating realistic
  synaptic microphysiology using {MC}ell}, Computational Neuroscience:
  Realistic Modeling for Experimentalists (E.~Schutter, ed.), CRC Press, 2001,
  pp.~87--127.

\bibitem{Thomas:1995:NPD}
J.~Thomas, \emph{Numerical partial differential equations}, vol.~22, Springer
  Verlag, 1995.

\bibitem{Twomey:2007:SMR}
A.~Twomey, \emph{On the stochastic modelling of reaction-diffusion processes},
  {M.S}c. {T}hesis, University of Oxford, United Kingdom, September 2007.

\bibitem{vanKampen:2007:SPP}
N.~van Kampen, \emph{{S}tochastic {P}rocesses in {P}hysics and {C}hemistry},
  3rd ed., North-Holland, Amsterdam, 2007.

\bibitem{vanZon:2005:GFR}
J.~van Zon and P.~ten Wolde, \emph{Green's-function reaction dynamics: a
  particle-based approach for simulating biochemical networks in time and
  space}, Journal of Chemical Physics \textbf{123} (2005), 234910.

\bibitem{Zauderer:1983:PDE}
E.~Zauderer, \emph{{P}artial {D}ifferential {E}quations of {A}pplied
  {M}athematics}, John Wiley \& Sons, 1983.

\end{thebibliography}
\end{document}